\documentclass[12pt,preprint]{aastex}
\usepackage{graphicx}

\shorttitle{Stability Constraints on the GJ~581 System}
\shortauthors{David Joiner et al.}
\slugcomment{Submitted for publication in the Astrophysical Journal}

\usepackage{natbib}
\begin{document}

\title{A Consistent Orbital Stability Analysis for the GJ~581 System}
\author{
  David A. Joiner\altaffilmark{1},
  Cesar Sul  \altaffilmark{2},
  Diana Dragomir\altaffilmark{3} \altaffilmark{4},
  Stephen R. Kane\altaffilmark{5},
  Monika E. Kress \altaffilmark{2}
}
\email{djoiner@kean.edu}
\altaffiltext{1}{Center for Science, Technology, and Mathematics, Kean University, 1000 Morris Ave, Union, NJ 07083.}
\altaffiltext{2}{Department of Physics \& Astronomy, San Jose State University, San Jose, CA 95192}
\altaffiltext{3}{Las Cumbres Observatory Global Telescope Network,
  6740 Cortona Dr., Suite 102, Goleta, CA 93117, USA}
\altaffiltext{4}{Department of Physics, Broida Hall, UC Santa Barbara, CA, USA}
\altaffiltext{5}{Department of Physics \& Astronomy, San Francisco State University, 1600 Holloway Avenue, San Francisco, CA 94132}

\begin{abstract}

We apply a combination of N-body modeling techniques and automated data fitting with Monte Carlo Markov Chain uncertainty analysis of Keplerian orbital models to radial velocity data to determine long term stability of the planetary system GJ~581. We find that while there are stability concerns with the 4-planet model as published by \cite{forveille2011}, when uncertainties in the system are accounted for, particularly stellar jitter, the hypothesis that the 4-planet model is gravitationally unstable is not statistically significant. Additionally, the system including proposed planet g by \cite{vogt2012} also shows some stability concerns when eccentricities are allowed to float in the orbital fit, yet when uncertainties are included in the analysis the system including planet g also can not be proven to be unstable. We present revised reduced chi-squared values for Keplerian astrocentric orbital fits assuming 4-planet and 5-planet models for GJ~581 under the condition that best fits must be stable, and find no distinguishable difference by including planet g in the model. Additionally we present revised orbital element estimates for each assuming uncertainties due to stellar jitter under the constraint of the system being gravitationally stable.

\end{abstract}

\keywords{methods: N-body simulations -- methods: numerical -- methods:statistical -- planetary systems -- stars: individual (GJ~581) -- -- technique: radial-velocity}

\section{Introduction}

In the last few years, results from the {\it Kepler} mission and radial velocity surveys have shown that small planets are significantly more common than giant planets. We are gradually discovering the properties and occurrence rates of the former population. Of particular interest are planets around M dwarfs because they contribute to our knowledge of planetary systems beyond those orbiting solar-type stars. The proximity of the habitable zone to the host star and the low stellar masses significantly facilitate the detection of lower mass, potentially habitable planets with the radial velocity method. 

GJ~581 is the first multiple system with all known planets having minimum masses smaller than that of Neptune, and its history as portrayed by the extensive publication record is rather tumultuous. This is in part due to the fact that although the system was discovered over eight years ago, the masses and orbital properties of some of the GJ~581 planets are still not well determined.

The first detection of a planet orbiting GJ~581 with a period of 5.4 days (GJ~581~b) dates to 2005 \citep{bonfils2005}. Subsequent studies revealed planets with periods of 12.9 days (GJ~581~c) and 83.4 days (GJ~581~d, later revised to 66 days \citep{udry2007}) as well as a lower mass planet at 3.15 days (GJ~581~e; \citep{mayor2009}). Shortly after, \cite{vogt2010} reported the discovery of 2 additional planets at periods of 433 days (GJ~581~f) and 36.6 days (GJ~581~g). The existence of planets f and g has been the subject of some debate. \cite{forveille2011} (hereafter F2011) submitted a paper including new HARPS data that disputes this finding. \cite{vogt2012} (hereafter V2012) have analyzed F2011's claims, and have found that the reduced chi-squared values as published by F2011 cannot be reproduced under assumptions of either astrocentric or Jacobi coordinates without removal of at least 5 high-residual data points, and that the orbital properties as published by F2011 are dynamically unstable on a timescale of hundreds of years, due to the high reported eccentricity of planet e. V2012 claimed that due to the uncertainty in the determination of the eccentricities for this system, it was better modeled with circular orbits to address the issue of dynamical stability. Their justification is the principle that finding a lower overall reduced chi-squared value with fewer total fit parameters was favored by the principle of parsimony. Their approach also allows probing for a possible planet masked by an artificially high eccentricity detection of planet d.  \cite{kress2013} found that relaxing the eccentricity of planet e to a lower yet still significant value could bring the system into dynamical stability. These findings were also reported by \cite{toth2013}, who studied key chaotic indicators and constrained the eccentricity of planet e to less than $0.2$. They also found that a 5 planet system including planet g could not be ruled out on the basis of dynamical stability. Additionally, \cite{gregory2011} applied a maximum likelihood method to the GJ~581 system and claims significant detection of a planet near $400$ days, but does not find significance for a planet near $33$ days.

We carry out an analysis which combines modeling the published radial velocities for this system and N-body integrations, in order to take into account gravitational interactions between the planets. Our work examines the stability of the system with 4 and 5 planets as a function of eccentricity and orbital inclination (assuming all planets are co-planar) over a timescale of 10 Myr, whereas \citeauthor{toth2013} integrate their models over only 1 Myr. \citeauthor{mayor2009} find that, for some initial conditions, the system becomes unstable (with planet e escaping) even after a few Myr, further justifying our choice of a longer integration timescale.  

We describe the N-body simulations and our method for fitting the radial velocities in Section 2.
Section 3 contains a discussion of computational issues.
We report our results in Section 4.
In Section 5 we address the impact on our results of including system inclination in our simulations.
We discuss our results in the context of the habitable zone of the system in Section 6, and conclude in Section 7.

\section{Method}

Common practice in determining the orbital properties and number of planets in an extrasolar planetary system involves fitting the radial velocity of a star with a Keplerian orbital model. This approach is well suited to both the use of heuristic non-linear stochastic minimization schemes, such as genetic algorithms and simulated annealing, and can also be coupled with Bayesian methods such as Monte-Carlo Markov Chain to produce a detailed analysis of uncertainties within fit models and relative likelihoods between models \citep{ford2005,ford2006}. While the use of a Keplerian model is efficient, and accurate in the case where orbits do not strongly interact, Keplerian orbits do not take into account interplanetary gravitational effects. This can be addressed to some degree by performing an n-body integration to determine orbital properties as opposed to using a Keplerian orbital fit; however in many cases the RV data available for a given star extends over thousands of days, whereas the timescale for instability may be anywhere from hundreds of years to millions of years. For that reason, typically exoplanetary orbital properties are validated through the use of longer term N-body models. These planetary systems must be modeled with high precision as current techniques using radial velocity data favor finding planets in close orbits around low mass stars, and orbits on the order of days for the innermost planet in a system are not uncommon, with a period as short as 0.74 days reported for 55 Cnc e \citep{endl2012}, and an innermost planetary orbital period of just over 3 days for GJ~581~e. The N-body models used must allow for high precision over long runtimes, while being practical to solve on modern computer systems. Symplectic integrators are a class of energy conserving integration schemes for canonical variables in generalized coordinate systems, and are well-suited to the types of orbital dynamics found in planetary systems. Many symplectic and hybrid-symplectic integration codes exist for orbital analysis. The simulations here were run using Mercury with a time step of 0.1 days and an accuracy parameter of $1.0 \times 10^{-12}$ \citep{chambers1999}.

The radial velocity is assumed to be the sum of the effects of $N$ non-interacting orbiting objects, with the parameters for the $i$th orbit defined by the semi-major amplitude ($K_i$), the mean anomaly at the initial time ($M_i$) used to define the true anomaly $f_i$ at future times in the model, the orbital period $T_i$, the eccentricity $e_i$, and the argument of periastron $\omega_i$. From \citep{beauge2007}

\begin{equation}
V_r(t) = \sum_{i=1}^{N}{K_i \left[ \cos(f_i +\omega_i) + e_i \cos(\omega_i)  \right]} + V_{r0}
\end{equation}

\begin{equation}
K_i=\frac{m_i \sin(I_i)}{m_{total}} \frac{2 \pi a_i}{T_i \sqrt{1-e_i^2}}
\end{equation}

\begin{equation}
m_{total}=\sum_{i=0}^N m_i
\end{equation}

where $V_{r0}$ is the relative velocity of the system independent of orbiting planets, $m_i$ is the mass of each planet or the central star in the case of $m_0$, and $a_i$ is the semi-major axis of the planetary orbit. 

The mean anomaly $M_i$ can be related to the true anomaly via the eccentric anomaly $u$ as

\begin{equation}
\tan(f_i/2) = \sqrt{\frac{1+e_i}{1-e_i}} \tan(u_i/2)
\end{equation}

\begin{equation}
u_i - e_i \sin(u_i)=\frac{2 \pi}{T_i}(t-\tau_i)
\end{equation}

where $t-\tau_i$ is the time since periastron, and can be related to the mean anomaly at epoch as $M_i = (2 \pi / T_i ) ( t_0 - \tau_i )$.

In addition to the computational challenge of long run-times and short time-steps faced ($\sim 1/100$ of the period of the innermost planet) evolved for a significant fraction of the lifetime of a typical solar system (typically $10^7 years$), in the case of multi-planet systems these systems are potentially tightly packed and thus sensitive to the values of the orbital properties being predicted, particularly the eccentricity. This can be of concern in the case where RV fitting suggests a high eccentricity and questions abound whether that high eccentricity is masking the signal of an additional planet at half the period. 

The recent literature regarding GJ~581 has largely centered on the issue of the number of planets in the system, in particular the proposed planets GJ~581~f and GJ~581~g. f is a longer period planet compared to the rest of the system, and is not expected to interact with the inner planets in the system. Planet g is proposed to be situated at roughly half the orbital period of d (33 days compared to 66), with the controversy around its existence centering around whether planet d is a high eccentricity orbit, or a low eccentricity orbit that is masking planet g.

Our approach is to first reproduce the RV fits for a 4- and 5-planet model of GJ~581, with and without a planet between c and d. Our primary focus is to establish whether the stability of the system can be used to better determine the existence of planet g, and we do not include planet f in our 5-planet models of b,c,d,e, and g. Planet f does not strongly affect the stability of the system or the RV fit results of b-e, and as such is treated separately.  The existing HARPS data on GJ~581 is used as published in F2011. Fits are initially performed minimizing the reduced chi-squared value with zero stellar jitter $j$, given by

\begin{equation}
\chi_R^2 = \frac{1}{n_{data}-n_{fit}} \sum \frac{ (v_{model}(t_i)-v_{data \hspace{1 pt} i})^2}{\sigma_{data \hspace{1 pt} i}^2+j^2}
\end{equation}

where $n_{data}$ is the number of radial velocity data points, $n_{fit}$ is the total number of parameters used to fit the data (typically 5 orbital parameters per planet and one velocity offset), $v_{model}$ is the radial velocity predicted by the model, $v_{data}$ is the measured radial velocity, $\sigma_{data}$ is the uncertainty in the measured radial velocity, and $j$ is the stellar jitter. 

Our minimization scheme involves iterating between genetic algorithm and simulated annealing solutions, using multiple ensembles in parallel. Ensembles at each iteration that fail an F-test at a 95\% confidence level compared to the best solution are reset to the current best solution. To determine the posterior distribution of each fit parameter, each ensemble solution is then randomized and used as a separate starting point for an MCMC analysis. For the results presented here, runs were performed using a parallel stochastic minimization scheme that alternates between genetic algorithms (GA) and simulated annealing (SA), with 160 ensembles per fit. Each ensemble iterated 5 times through a GA minimization followed by an SA minimization, with an MCMC error estimate applied to the fit results. For the GA minimization, 100 generations with a population size of 1000 were used. The 5 best fits were kept each generation and allowed to recombine to create the next generation, fit parameters were mapped to a real number line and recombination occurred via use of normal random deviates around current fit values. The SA minimization was performed with a cooling factor of $0.999$ and a maximum number of iterations of $50000$ trials per iteration, with $200$ trials per temperature readjustment. The MCMC analysis was performed for 500000 steps per ensemble with a burn-in of 100000 steps, and 10000 steps were randomly sampled across all ensembles for statistical analysis. Typical run-times were on the order of 10 minutes of wall time on 20 cluster nodes with 8 cores each, or about 1 CPU day.

After the initial fit, the stellar jitter required to bring the reduced chi-squared value to 1 is determined using a bracketing algorithm. RV fits are then re-run with jitter included in the reduced chi squared calculation during minimization, and an analysis of uncertainty of fit parameters is performed using a Monte-Carlo Markov Chain approach with a likelihood of $e^{-\chi_R^2 / 2}$. We use eccentricity relative to a cutoff or N-body model time until collision/completion as discrepancy variables for the purpose of defining a p-value from the posterior distribution relative to a null hypothesis that the model fit is stable. While not a p-value in the traditional frequentist predictive sense (these p-values are relative to the posterior distribution, not necessarily to the model or the data directly), such p-values can still be used to indicate whether there is a discrepancy between the model and the data--particularly aspects of the data not well described by the model, with p-values very close to zero indicating a lack of overlap between the model and data \citep{gelman96}.

The posterior distribution of the fit parameters is used to determine a range of input values for a Monte-Carlo parameter space study of both the 4- and 5-planet models, with the attempt to try to determine stability islands which can be characterized by critical values of key model parameters. Typical run times for 1000 N-body simulations run in parallel on 100 8-core nodes were on the order of 2 days of wall time, or about 5 CPU years.

The posterior distributions from MCMC analysis are then recomputed based on constraints from those critical values, and used to estimate orbital properties with uncertainty in a manner that is consistent both with model fit and with system stability. Additionally, an RV fit without jitter but with stability constraints is performed to calculate a reduced chi-squared value for the system given that the system should be stable. This stable, zero-jitter reduced chi-squared value is then used to compare the 4- and 5-planet fits to determine the significance, if any, of the detection of planet g.

Simply limiting the search space in the RV fitting methods can result in an overestimation of the portion of the posterior distribution that meets the constraint of stability. There is some uncertainty in determining the specific value of eccentricities that result in unstable orbits, particularly in multi-planet models where one orbit may be stable with a high eccentricity but two nearby orbits with high eccentricity might not be. To validate the approach of using a critical value of key parameters to constrain the MCMC analysis, the results of the MCMC analysis without constraints but with jitter are additionally sampled as input to a set of N-body simulations, and a probability distribution of model parameters conditional upon both fitting the data and representing a stable system is constructed and compared to prior results.

\noindent\textbf{Workflow used for Radial Velocity and Stability Analysis}
\begin{enumerate}
\item Run RV-Keplerian-Astrocentric fit without jitter, use to estimate jitter
\item Run parameter space study of N-body models to $10^7$ years. For this round of N-body simulations, inputs are chosen over a uniform distribution covering the range of expected orbital parameters. Eccentricities are sampled from zero to twice their expected value.
\begin{itemize}
\item Check for what parameters most affect stability--eccentricity, mean anomaly, periastron, inclination (also parameters least constrained by RV fit).
\item Determine islands of stability, what are their characteristics?
\item Cutoffs for key variables (e.g. eccentricity of planet GJ~581~e) may be illustrated by graphing n-body model survival time versus parameter value or by examining histograms of stable and unstable model parameters
\end{itemize}
\item Run RV-Keplerian-Astrocentric fit with jitter but without parameter constraints, determine the fraction of posterior distribution, given estimates of uncertainty, that fall in stable regime. 
\item Run RV fits with jitter and parameter constraints to determine best estimate of parameter values consistent with stable system.
\item Sample MCMC from unconstrained RV fits with jitter and use as input to N-body run to validate constrained posterior distribution estimates. Reconstruct posterior distributions from stable N-body simulation inputs.
\item Repeat process for competing models of system.
\begin{itemize}
\item Compare competing systems using reduced chi square without jitter but constrained by stability, perform F-test.
\item Compare stable fractions of posterior distribution.
\end{itemize}
\end{enumerate}

\section{Computational Issues}

The primary computational needs of this approach lies in the solution of many parallel small-N N-body systems. Runtimes on a 2.6GHz CPU for a single attempt at evaluating stability using the symplectic integration code Mercury with a timestep of $0.01$ and an accuracy parameter of $10^{-12}$ are on the order of $1-2$ days, though this time may be less in the event of an unstable system as integration can be stopped if close encounters, collisions, or ejections occur. Due to the small value of $N$ when studying exoplanetary systems, there is little, if any, usefulness in parallelizing a single stability test; however as there can be a large uncertainty for a given orbital parameter it is useful to run many different stability tests in parallel. Our typical Monte-Carlo parameter space studies range from 1000-10000 runs depending on the number of parameters that potentially affect stability, with an additional 1000 runs of the posterior distribution of the 4- and 5-planet MCMC analyses. These are run on a 130-node, 1040-core CPU cluster, with one core per simulation. Typical run-time per stability analysis assuming 1000 models is on the order of a couple of days. Jobs are created through a series of PERL scrips and stored in individual directories, and are scheduled using a PBS scheduler, and an allocation of roughly 5 CPU years is required per stability analysis. Analyses with greater than 1000 runs typically were performed only when a large number of scenarios tested led to instability and greater detail was thus required--this benefited in terms of CPU requirements by the early termination of many runs and did not require significantly greater allocated time.

While individual N-body simulations are not meaningfully parallelizable for this problem due to the small value of N, the number of independent simulations that are performed allow for parallelism across the ensemble. Much of this will be suitable to performance on General Purpose Graphics Computing Unit (GPGPU) hardware, and a prototype code has been developed to perform symplectic integration on many parallel N-body simulations with the purpose of flagging each run as either stable or unstable. Testing of this prototype code on an Nvidia Tesla K20c indicates that a similar analysis could be performed on the Tesla card in roughly 3 times the total wall time as on our cluster, but for a significant reduction in hardware and power costs. This approach to stability analysis for exoplanetary systems will be well suited to a small dedicated GPGPU workstation, and larger scale application of this approach to many systems could be enabled on a modest GPGPU cluster.

\section{Results}

\subsection{4-planet model}

\begin{figure*}
\begin{tabular}{ll}
\includegraphics[width=3.25in]{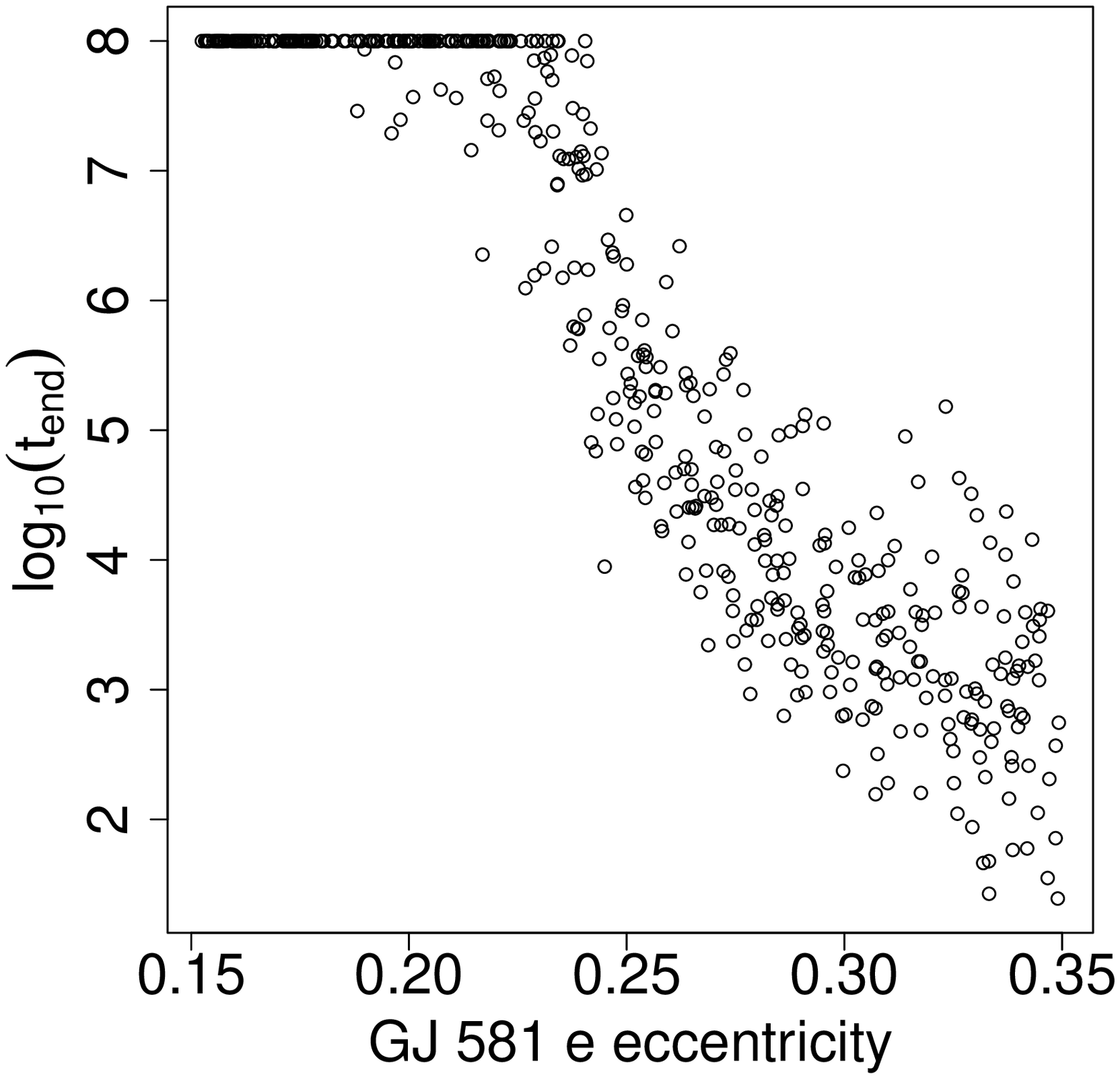}
&
\includegraphics[width=3.25in]{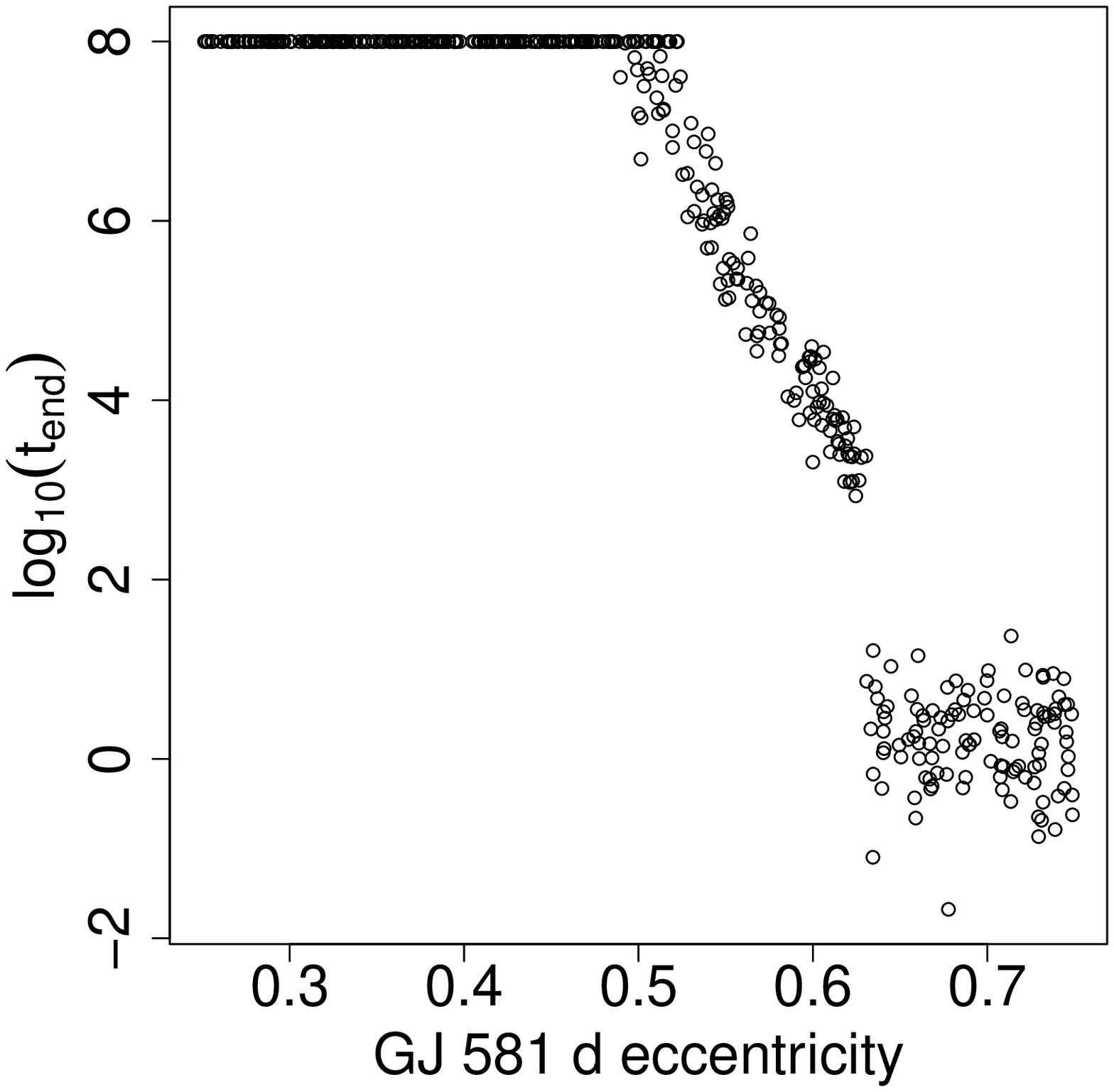}
\\
\end{tabular}
\caption{Model time versus either completion or collision versus eccentricity for 4-planet simulations of GJ~581. Model parameters are taken from F2011 except for phase variables. The first plot shows the effect of varying the eccentricity of planet e in the parameter space exploration while leaving d fixed. The second allows d to vary while leaving e fixed. \label{fig:EDecc}}
\end{figure*}

The 4-planet model as published in \cite{forveille2011} for GJ~581 has two potential concerns with regards to stability, the non-negligible eccentricities of planet e ($e_e=0.31$) and planet d ($e_d=0.25$). In order to test the impact of these two parameters, N-body models were run in a Monte-Carlo parameter space study. The eccentricities of each planet were picked randomly from a uniform range. Additionally, the initial mean anomaly and the relative periastron between orbits b and e was chosen randomly from $0$ to $2 \pi$. All other values were fixed to the 4-planet best fit of F2011. Figure~\ref{fig:EDecc} shows the results for different values of e and d's eccentricities varied one at a time. Increasing the eccentricity of planet e beyond a value of about $0.2$ or the eccentricity of d beyond $0.5$ results in 4-planet models that are not stable to a timescale of $10^8$ years.This was not significantly affected by the relative periastron of the orbits. Note that the transition from stability to instability occurs over a fairly small range of eccentricities in both cases, and that for increasing eccentricity the lifetime of the systems until collision falls off rapidly, with little difference between the cutoffs for stability to $10^8$ years compared to $10^7$ years. For computational purposes, remaining models are run to $10^7$ years.

The Keplerian RV fit for a 4-planet model is shown in Table~\ref{tab:4planetnoj}, and agree well with the values presented in F2011 with the exception of the RMS and $\chi_{R}$ values. V2012 has noted this discrepancy and linked it to the possibility that data points with the largest residuals may have been excluded in F2011's analysis. 

\begin{figure*}
\begin{tabular}{ll}
\includegraphics[width=3.25in]{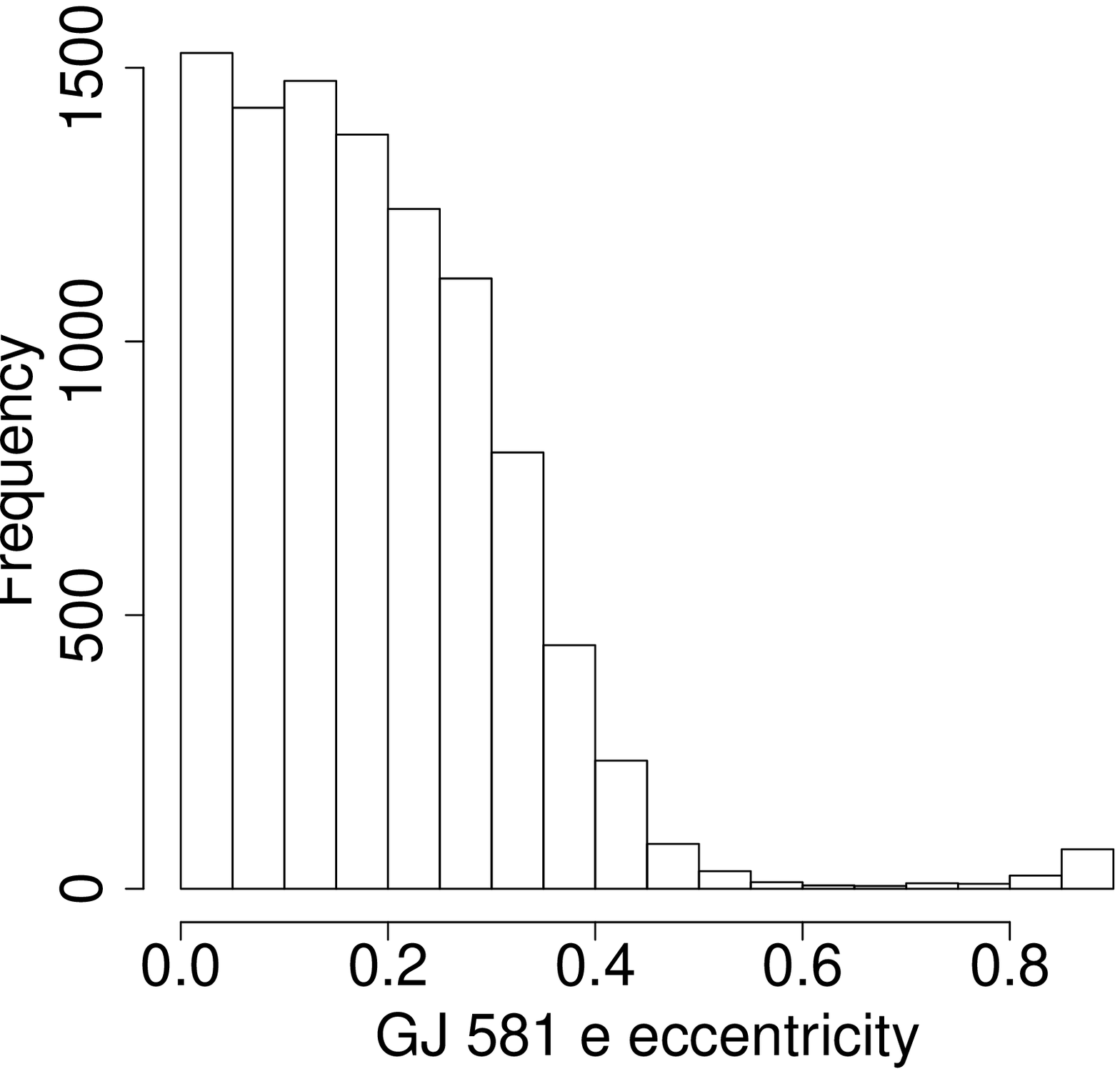}
&
\includegraphics[width=3.25in]{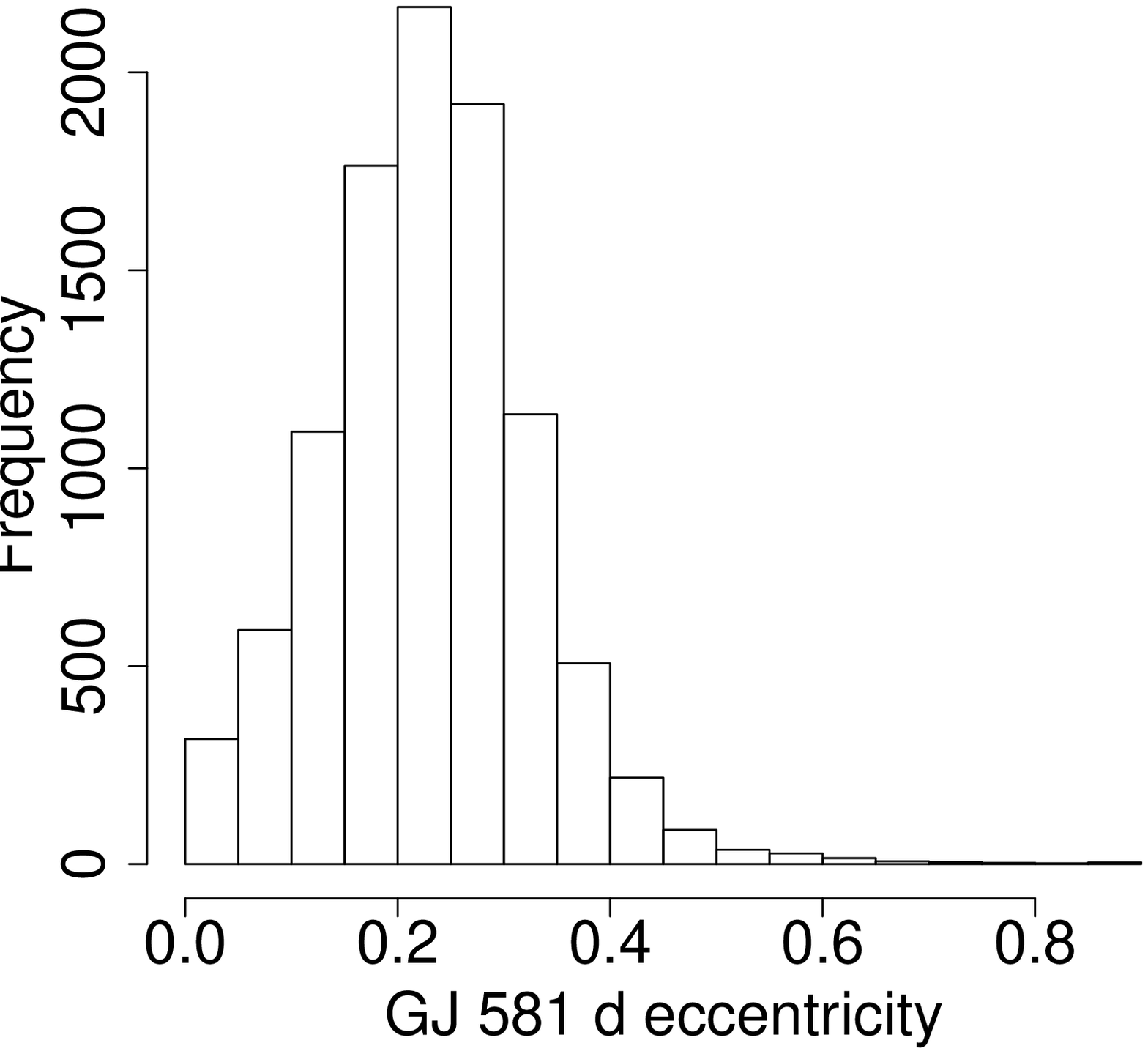}
\\
\end{tabular}
\caption{Monte-Carlo Markov Chain Posterior Distribution of the eccentricity of planets e and d in a 4-planet Keplerian model of GJ~581 which includes jitter but is not artificially constrained based on stability concerns. $58.7 \%$ of the distribution of the eccentricity of planet e lies at an eccentricity below 0.2. $99.0\%$ of the posterior distribution of the eccentricity of d lies below a value of 0.5. \label{fig:EDecc_jitter_mcmc}}
\end{figure*}

Orbital elements for a 4-planet model with a jitter of 1.53 are shown in Table~\ref{tab:4planet}. Figure~\ref{fig:EDecc_jitter_mcmc} shows the posterior distribution of the eccentricities of planets e and d. The portion of the distribution for which the eccentricity of e is less than $0.2$ is $58.7 \%$. Multiple studies have criticized the F2011 solution as being dynamically unstable, yet our results suggest that a hypothesis that the 4 planet solution is unstable fails a test of statistical significance.

The posterior distributions were then limited under the constraints of requiring the eccentricity of planet e being less than 0.2 and the eccentricity of planet d being less than 0.5, based on stability requirements. A stellar jitter of 1.53 is assumed. See Table~\ref{tab:4planetc}.

\subsection{5-planet model}

\begin{figure}
\plotone{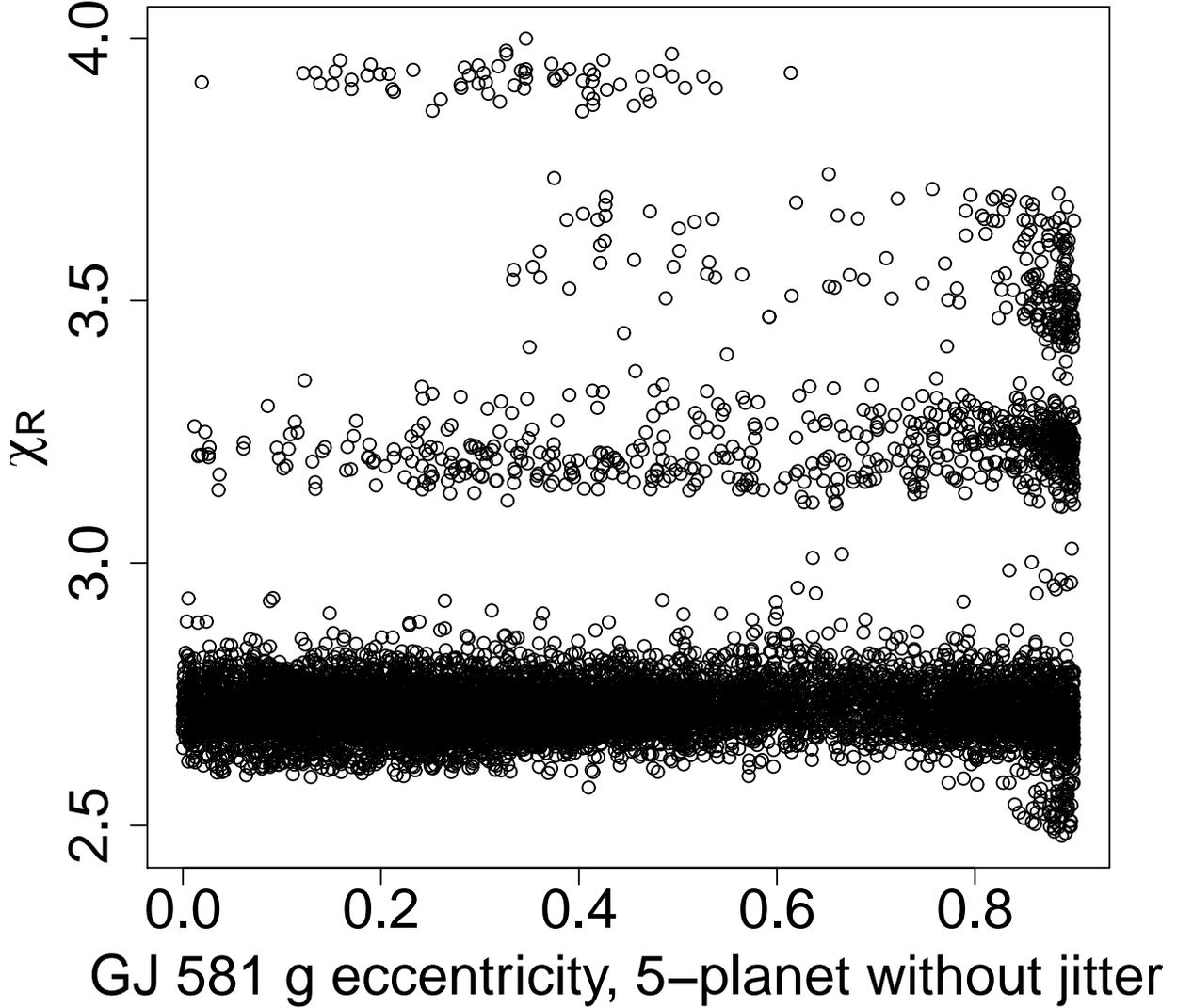}
\caption{Posterior distribution of the eccentricity of planet g in a 5-planet model, plotted versus $\chi_R$. Jitter is not explicitly included in the calculation of $\chi_R$ and no artificial constraints are placed on fit values for stability purposes. Note that while a solution exists with a global minimum $\chi_R$ the eccentricity of the solution is unphysically high, thus the value of $\chi_R$ for the 5-planet fit should be taken assuming lower eccentricities for planet g. Note that four similar local minima can be seen as clusters in this figure. The local minima with $\chi_R<3.5$ fall within a $99\%$ confidence interval.   \label{fig:Gecc_chired_5_nojitter}}
\end{figure}

Looking at a 5-planet Keplerian model to test for the signature of proposed planet g (see V2010 and V2012) in Table~\ref{tab:5planetnoj}, one sees that there is a minimum solution with a reduced chi squared of 2.48; however this value occurs only for solutions with an unphysically high value of the eccentricity of planet g. This can be seen in Figure~\ref{fig:Gecc_chired_5_nojitter}. Overall, the primary effect of including planet g on the orbital properties of the other planets is that the eccentricity of planet d is reduced. Even after discarding results centering around a minimum with a clearly too high value of the eccentricity of g, there is still a large range of values that the eccentricity of g may take on while fitting the data.

\begin{figure*}
\begin{tabular}{ll}
\includegraphics[width=2.75in]{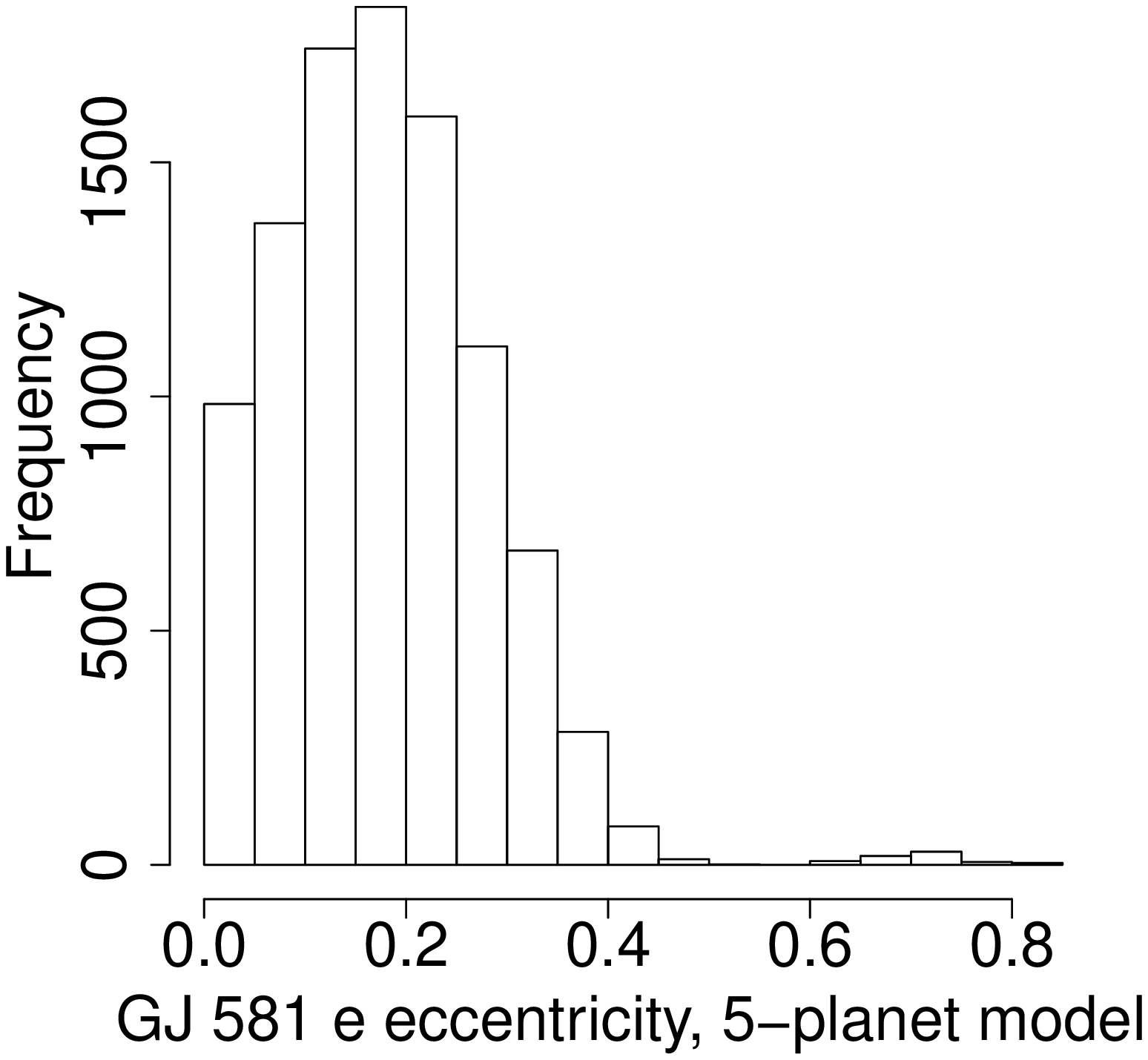}
&
\includegraphics[width=2.75in]{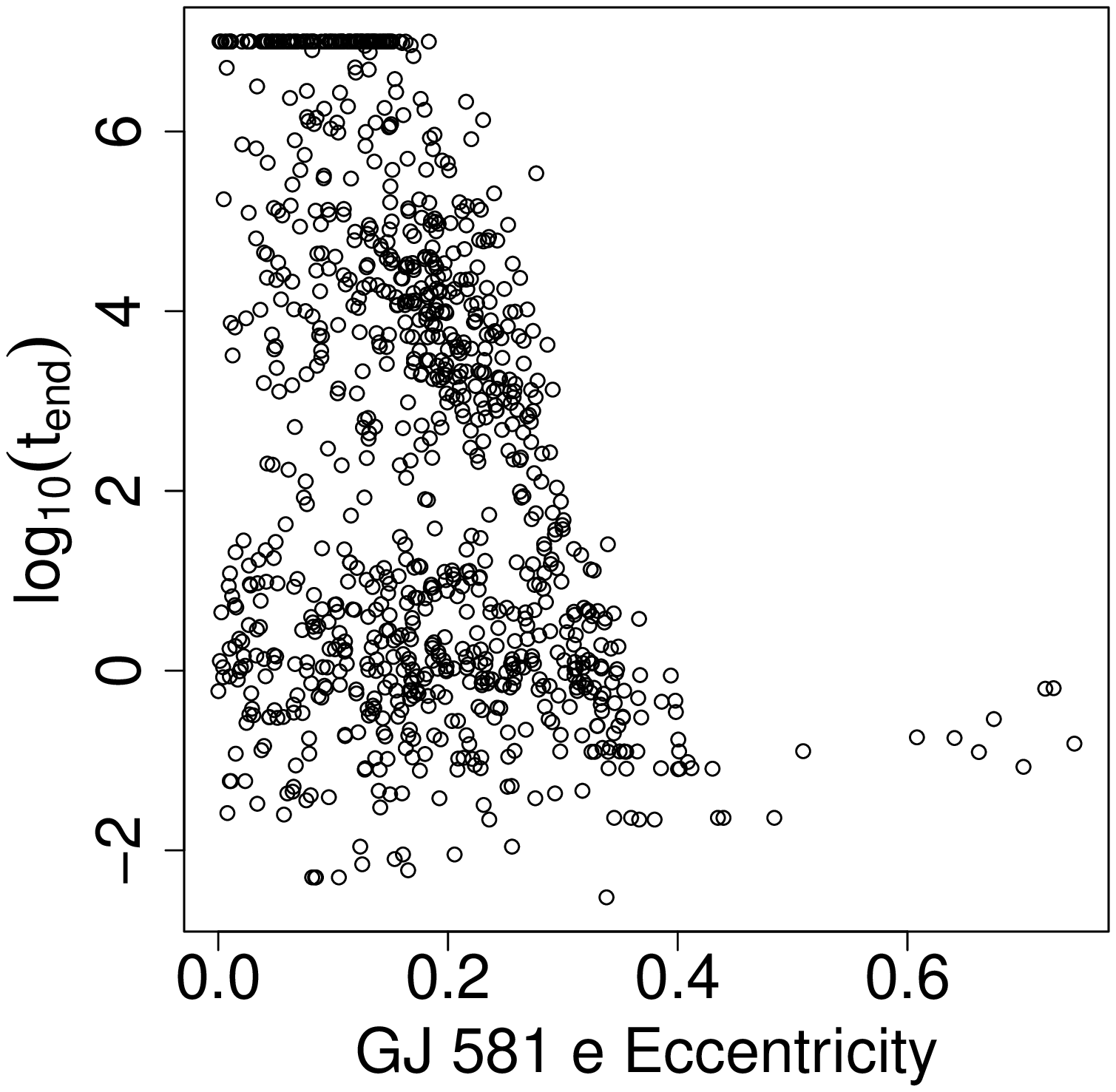}
\\ [-0.4in]
\includegraphics[width=2.75in]{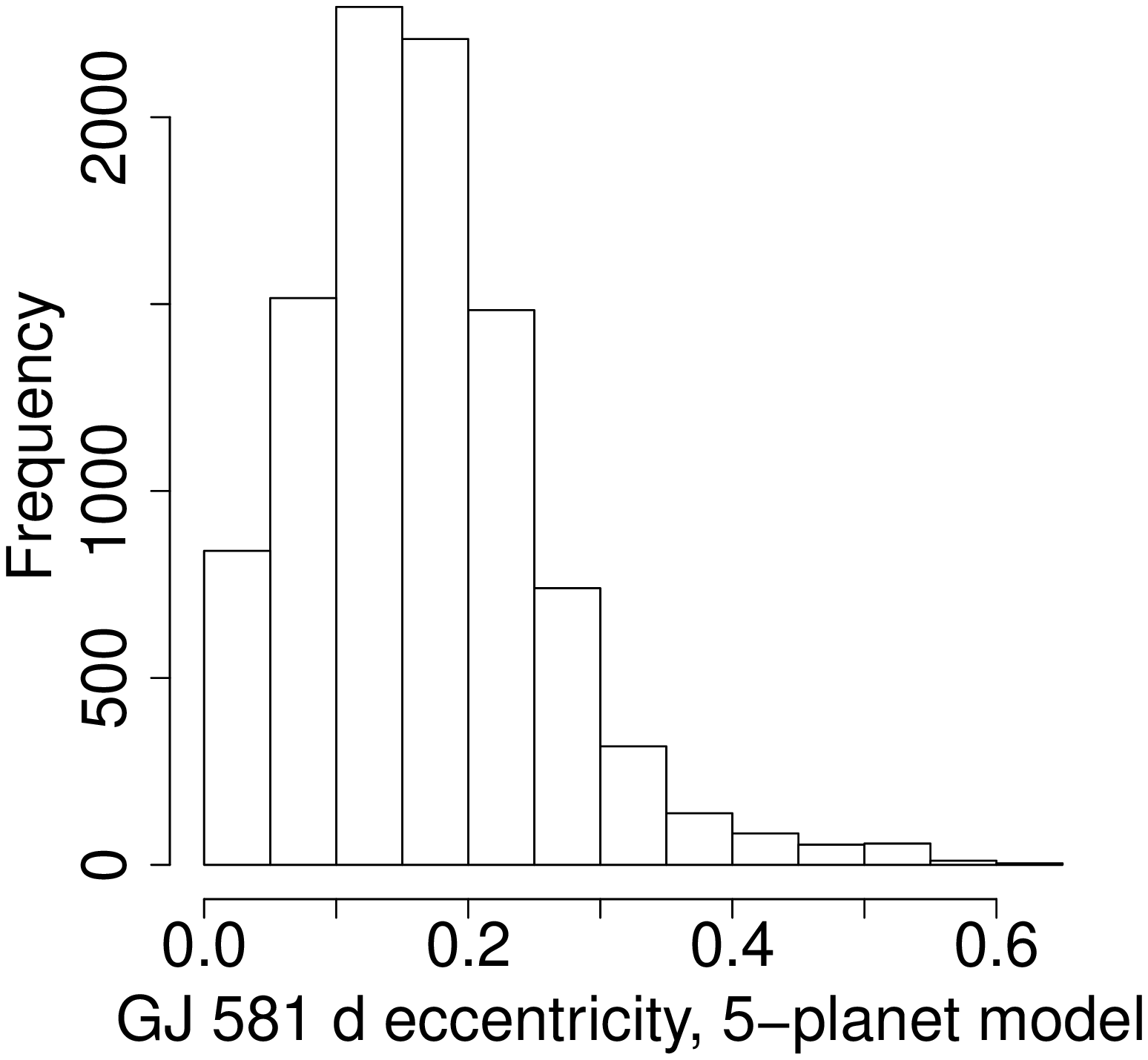}
&
\includegraphics[width=2.75in]{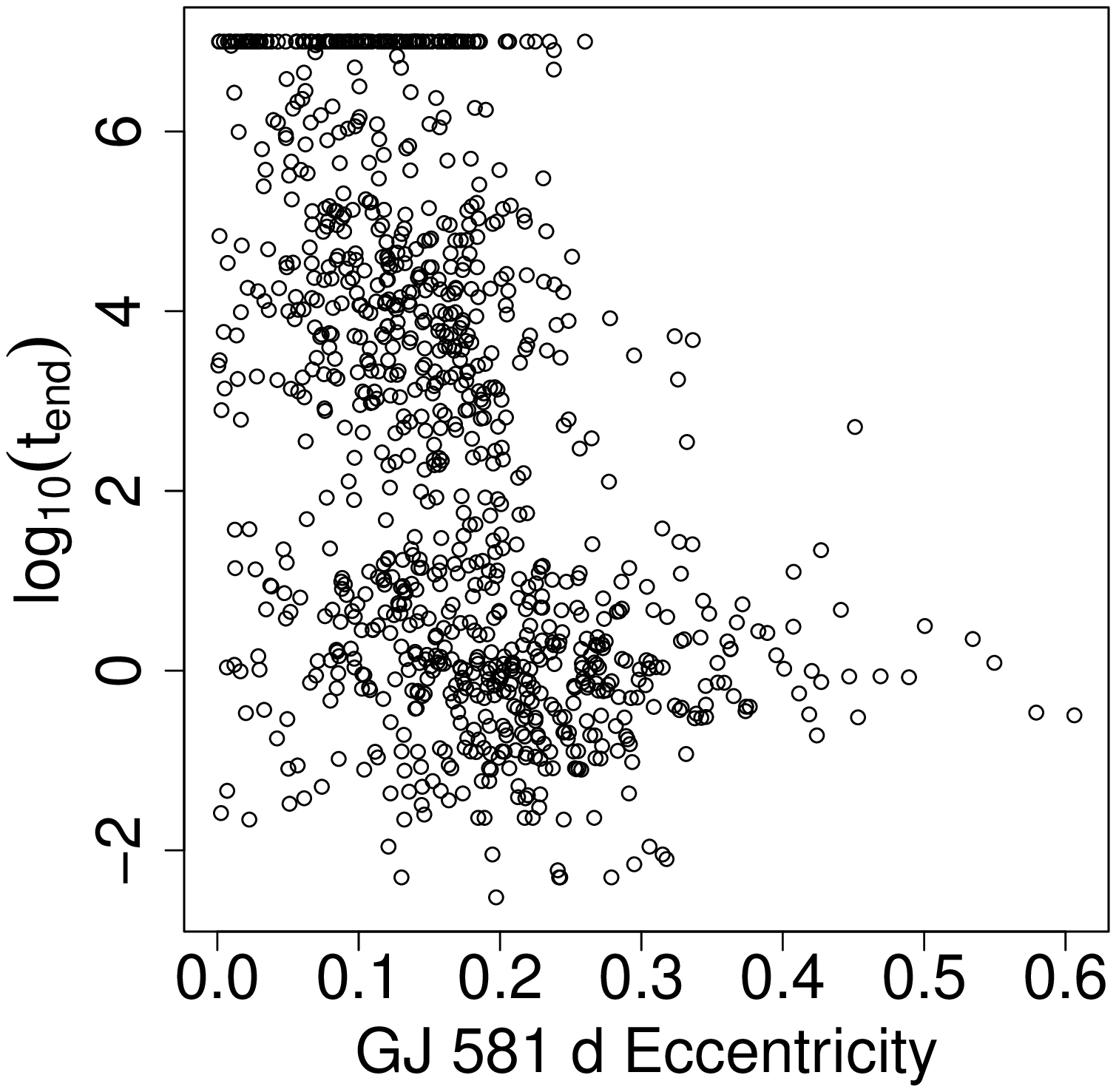}
\\ [-0.4in]
\includegraphics[width=2.75in]{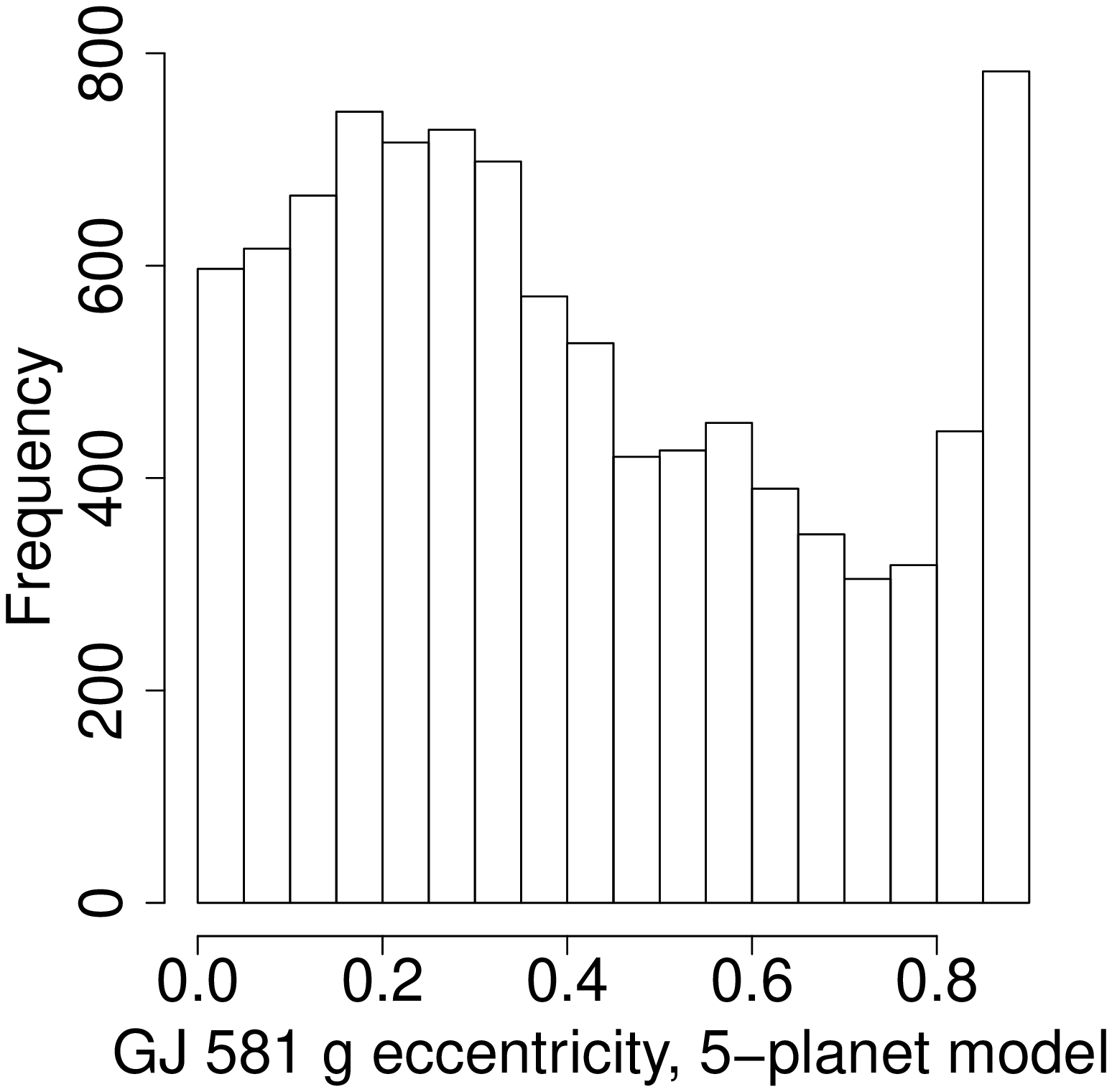}
&
\includegraphics[width=2.75in]{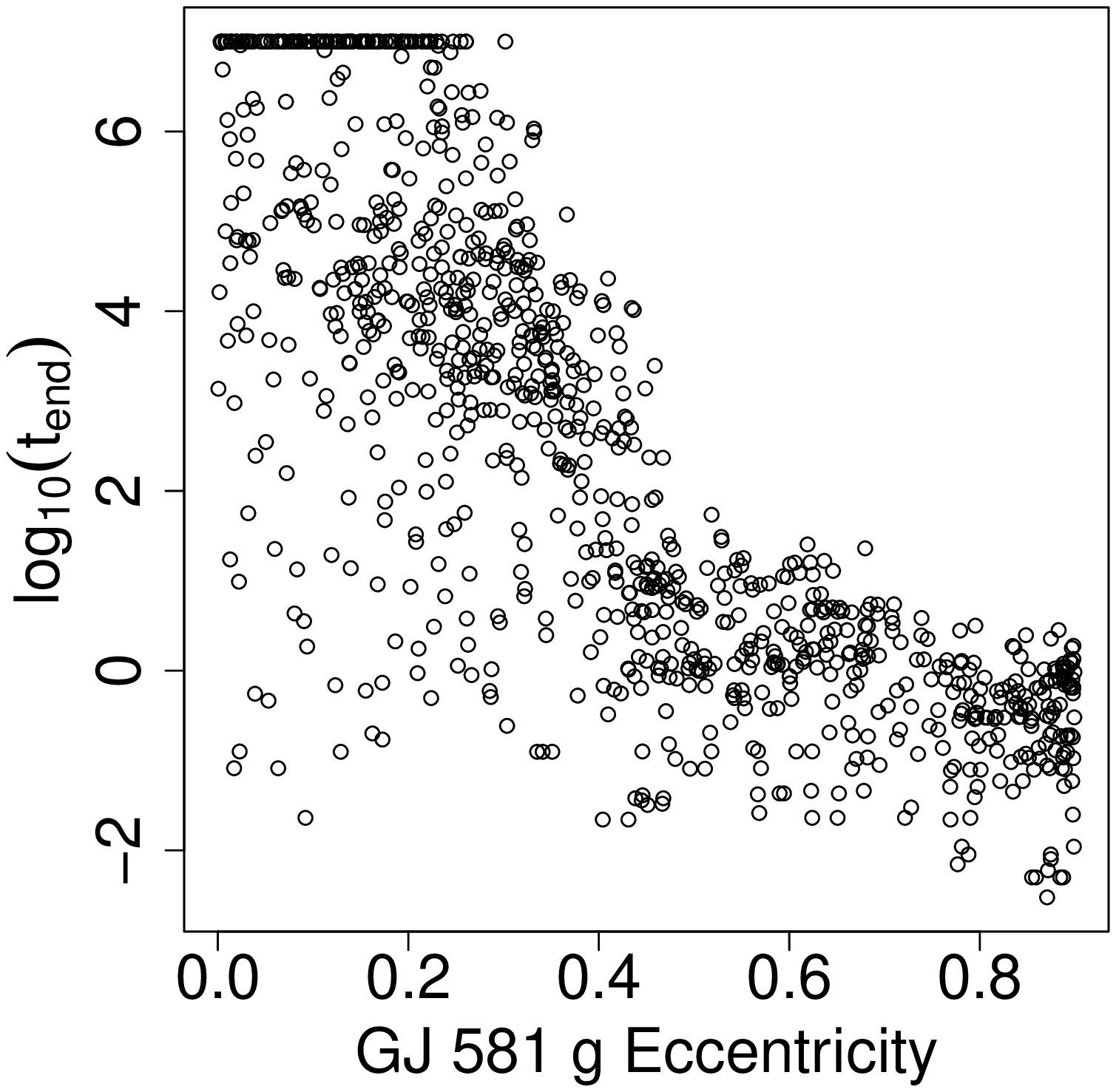}
\\ [0.3in]
\end{tabular}
\vspace{-0.5in}
\caption{Posterior distributions of the eccentricity of planets e, d, and g in a 5-planet model, with jitter explicitly included in $\chi_R$ calculations but with no artificial constraints added to enforce stability. Eccentricities are plotted both as histograms and as a scatter plot of eccentricity versus the lifetime of an N-body simulation before either completion at $10^7$ years or collision. 
\label{fig:EDGecc_mcmc_5}}
\end{figure*}

After including jitter in the 5-planet model RV fit, planet g still shows a wide range of eccentricities that fit the data with a large portion of the posterior distribution centered around unphysically high values--though to less of an extreme than the 5-planet RV fit without jitter. Results are summarized in Table~\ref{tab:5planet}. The posterior distributions of the eccentricities of planets e, d, and g are shown in Figure~\ref{fig:EDGecc_mcmc_5},along with the time until collision or completion of an N-body simulation based on each point in the posterior distribution. As in the 4 planet model, the eccentricity of planet e appears to be limited to no more than 0.2 for the 5 planet model. Planet d, similarly seems to have the eccentricity limited to no more than about 0.25 based on stability concerns. Planet g should not have an eccentricity more than 0.25. When compared to the posterior distributions of e and d, this represents about half of each distribution. For planet g, the posterior distribution is nearly uniform and does not seem to indicate a preference in the model for any particular eccentricity.

A RV fit to a 5-planet model with the eccentricity of e constrained to less than 0.2 and the eccentricity of g and d less than 0.25 without jitter has a best-fit reduced chi-squared of $2.58$ and a jitter of $1.46$. The fraction of the unconstrained posterior distribution that falls within these limits is $21\%$. An RV fit with 5 planets, jitter, and stability constraints is shown in Table~\ref{tab:5planetc}.

Comparing the best-fit, no-jitter, reduced chi-squared values of the 4- and 5-planet models, we see that the comparison is between a reduced chi-squared of 2.78 (4-planet) and 2.58 (5-planet). This represents an F-value of 1.078, with degrees of freedom of 219 and 214, or a p-value in an F-test of 0.29. This, combined with the relatively small portion of the 5-planet model posterior distribution that falls in a stable regime ($21\%$) and a value for the semi-major velocity of the planet signal of $0.89$ compared to a stellar jitter on the order of $1.4$ or higher, and an average instrumental error of $1.24 m/s$, does not suggest a statistically significant determination of the existence of planet g given the current data.

We similarly have tested a 5-planet model for GJ~581 including planet f near 400 days, but not including planet g. We find minima with periods near $400$ days and near $200$ days, both of which have posterior distributions for the eccentricity that skew unrealistically high, and when constrained to physical values do not find significant improvements in the $\chi_R$ values for the fit. ($\chi_R=2.69$ for a model with a planet at a period of $391$ days and $\chi_R=2.52$ for a planet at a period of $191$ days, provided we limit the results to a conservative upper limit on eccentricity of $0.5$. Neither $\chi_R$ value is significant using an F-test compared to the 4-planet best stable fit of $\chi_R=2.78$.) In neither case did including f in the RV fit significantly change the orbital parameters or uncertainties for planets b-e.

\subsection{Stability sampling of the posterior distribution}

\begin{figure*}
\begin{tabular}{ll}
\includegraphics[width=3.25in]{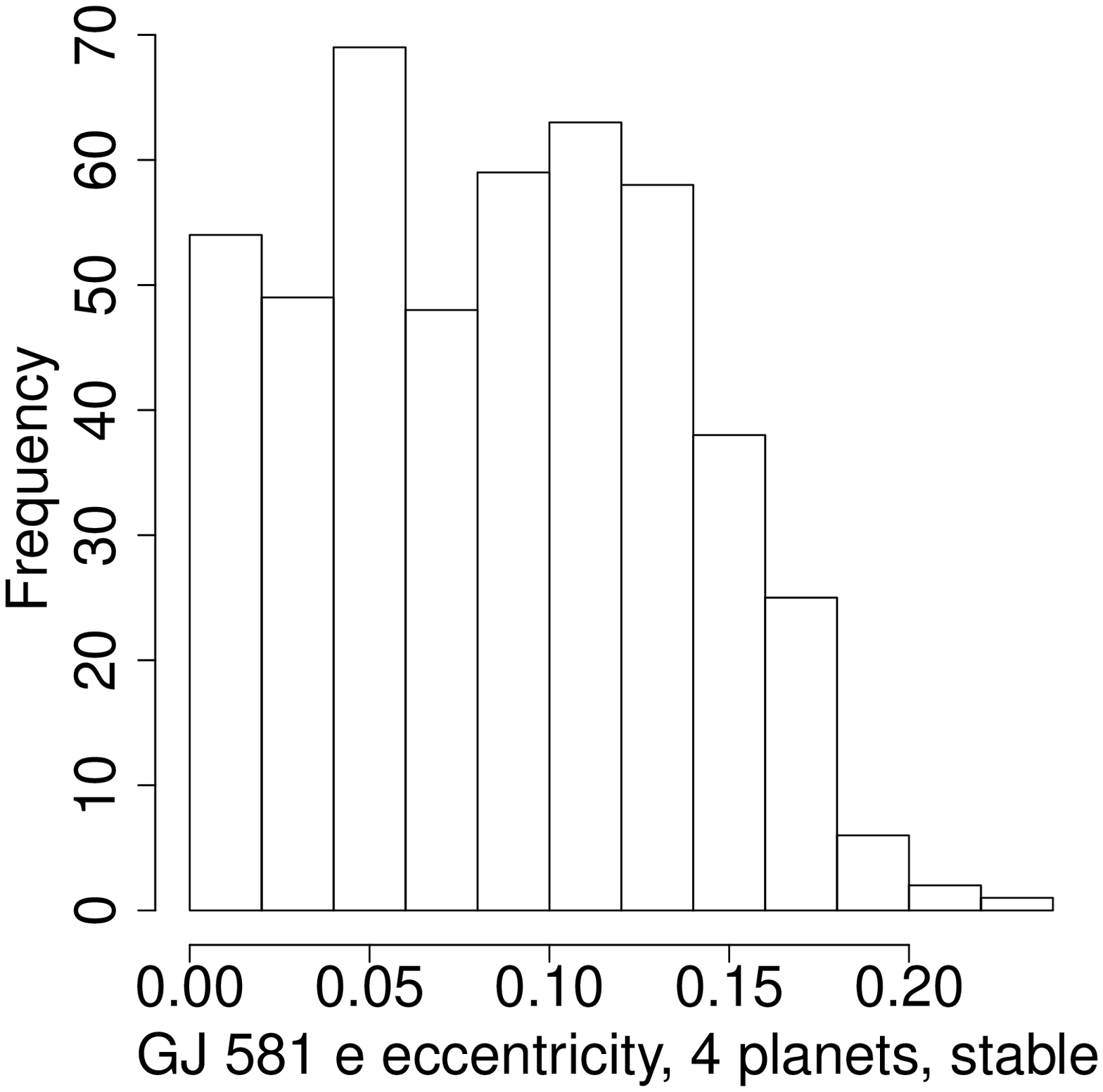}
&
\includegraphics[width=3.25in]{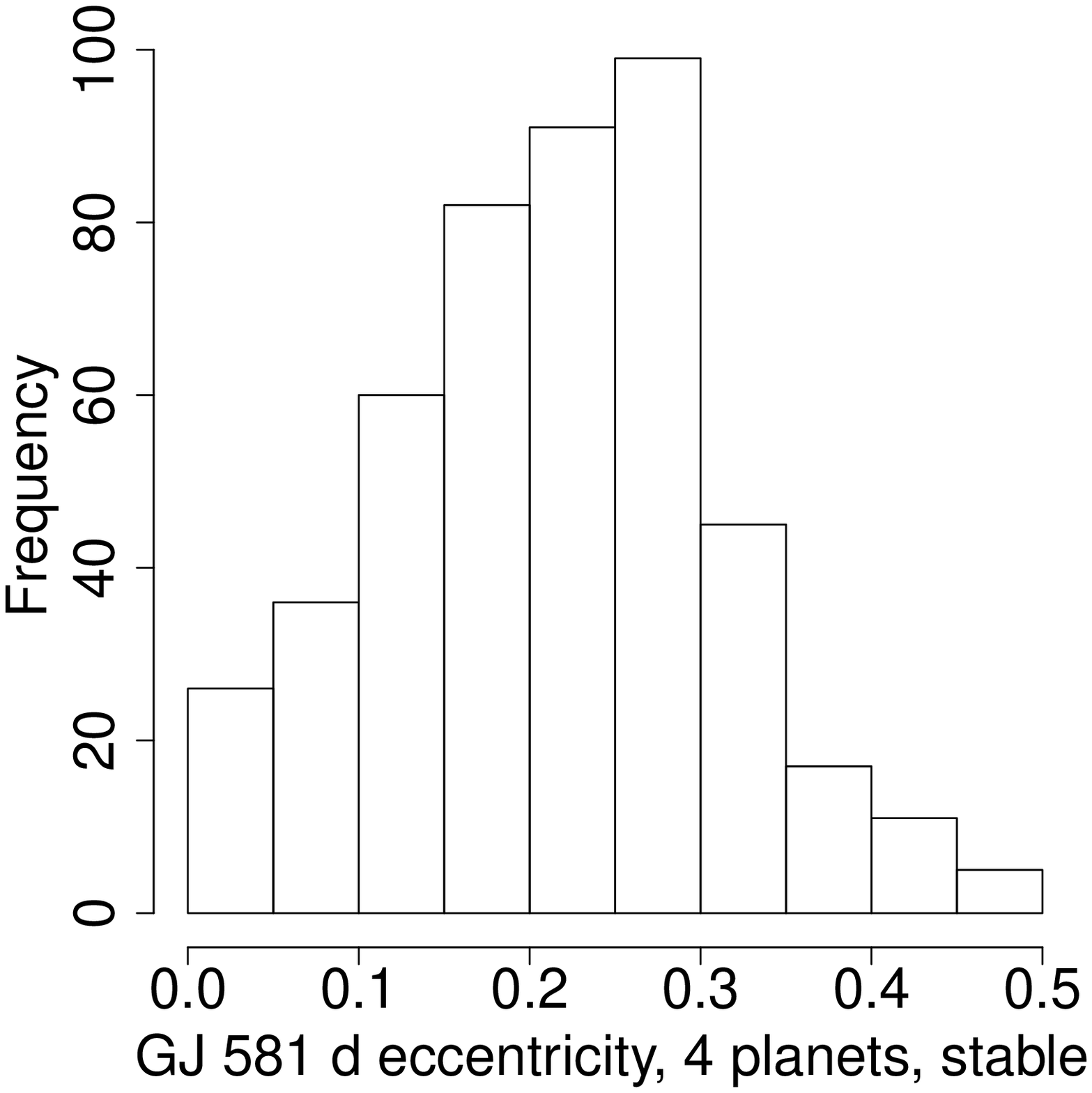}
\\
\end{tabular}
\caption{Posterior distributions of eccentricities of planets e and d in a 4-planet model conditional upon system being stable to $10^7 yrs$.\label{fig:EDecc_mcmc_stable_4}}
\end{figure*}

\begin{figure*}
\begin{tabular}{ll}
\includegraphics[width=3.25in]{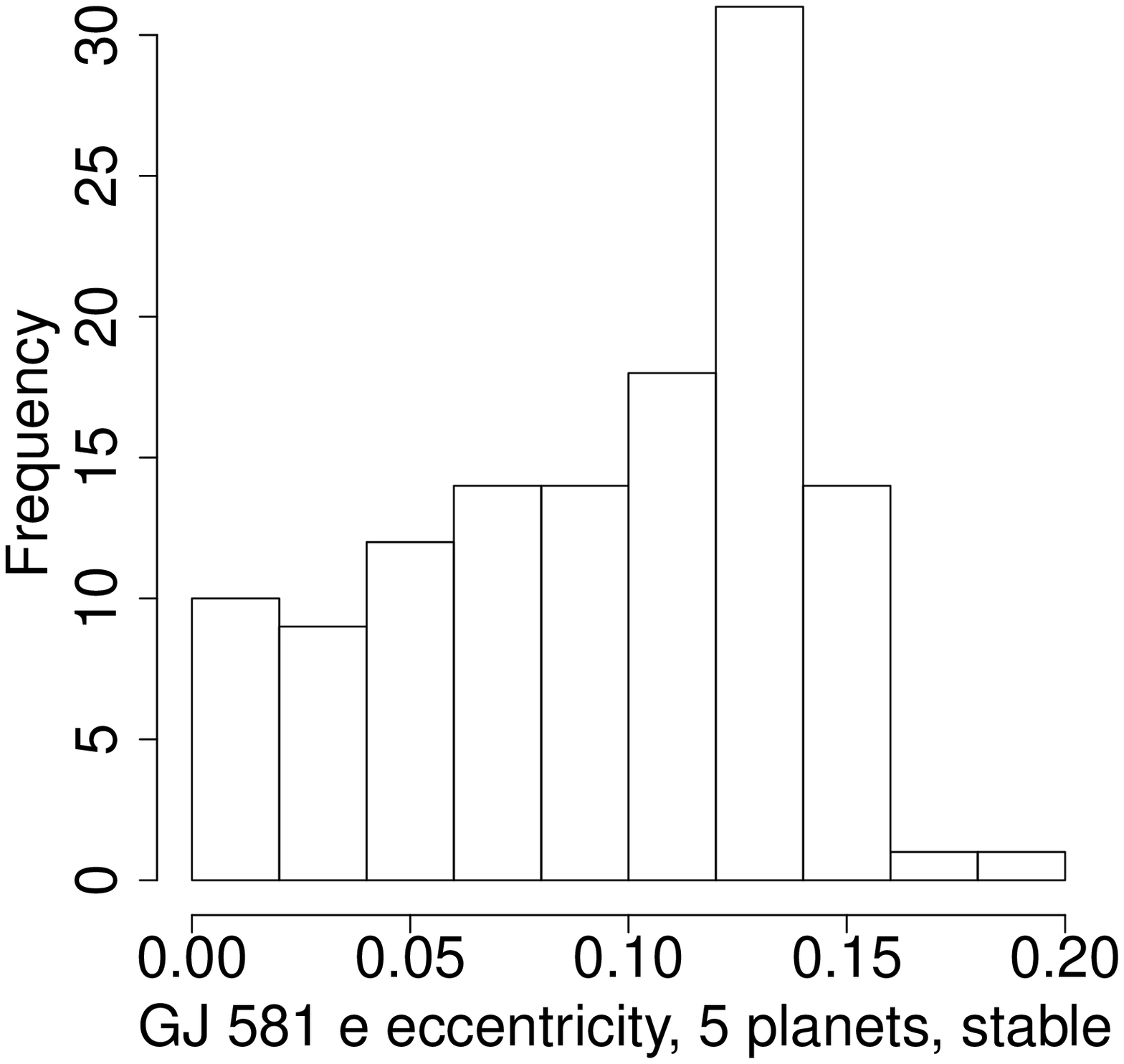}
&
\includegraphics[width=3.25in]{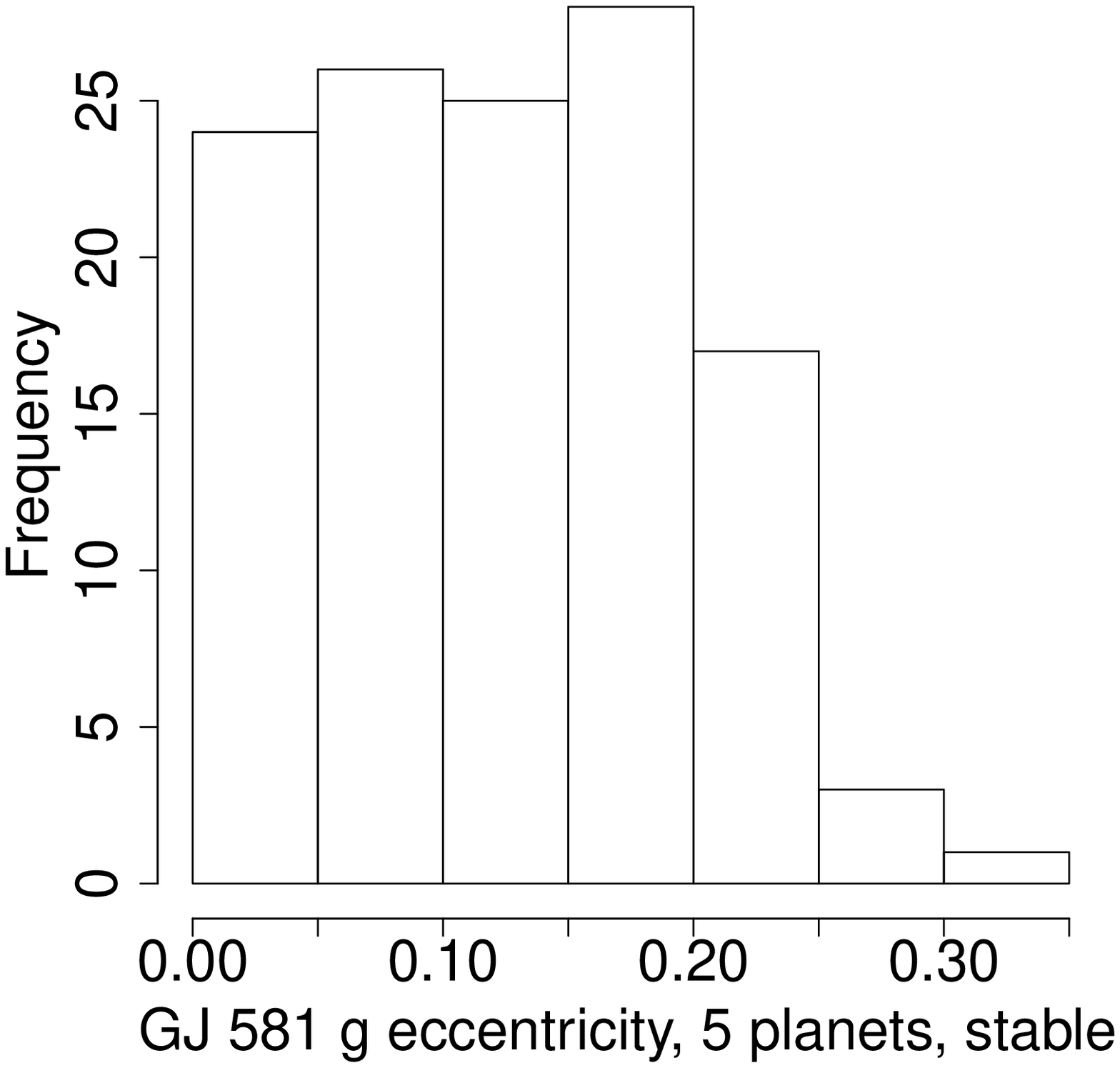}
\\
\includegraphics[width=3.25in]{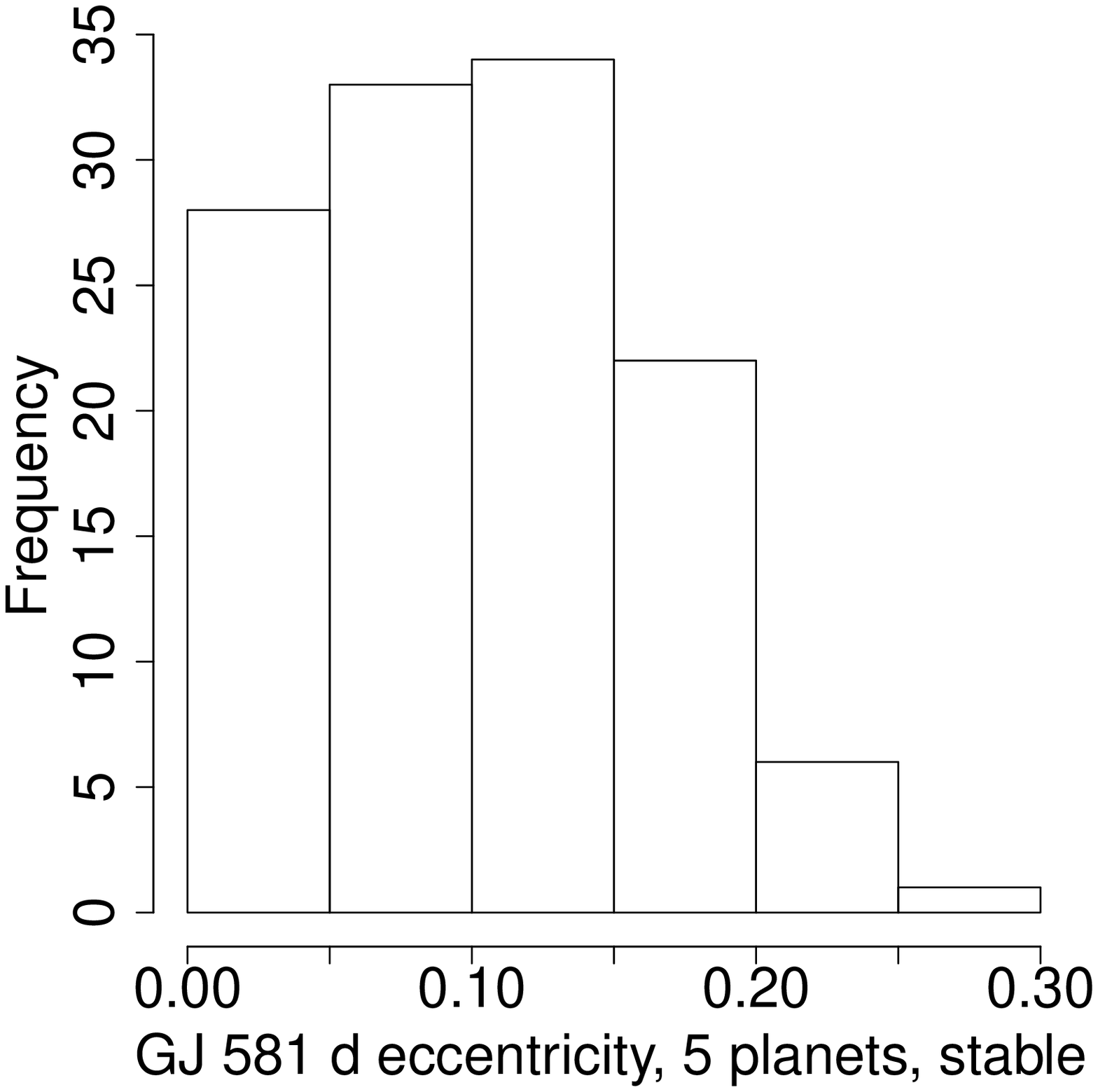}
& \\
\end{tabular}
\caption{Posterior distributions of eccentricities of planets e, g, and d in a 5-planet conditional upon system being stable to $10^7 yrs$.\label{fig:EDGecc_mcmc_stable_5}}

\end{figure*}

As a validation of the stability constraints placed on the eccentricities within the MCMC analysis, a sample of 1000 fit attempts was taken from the posterior distributions of the 4-planet and 5-planet models with jitter. The posterior distributions of the eccentricities of e (4-planet), d (4-planet), e (5-planet), g (5-planet), and d (5-planet) are shown in figures \ref{fig:EDecc_mcmc_stable_4} and
\ref{fig:EDGecc_mcmc_stable_5}. Note that the fraction of the posterior distribution that is stable using this method is less than that obtained by simply applying a cutoff to the maximum value of the eccentricities in the posterior distribution as using a simple cutoff can overestimate stability, neglecting the possibility that models with a high eccentricity for one planet may have more stringent restrictions as a result on the eccentricities of an interacting planet. We find that $12\%$ of the posterior distribution is stable for the 5-planet model RV fit, and $47\%$ for the 4 planet model. The fraction of the 5 planet posterior distribution sampled that fell in the stable regime may have been affected by the relative insensitivity of the Keplerian data fit to the eccentricity of planet g.

\section{Inclination}

A transit search has been carried out for planet e independently by \cite{dragomir2012b} and F2011. \citeauthor{dragomir2012b} found no transits within the 3$\sigma$ predicted transit window, ruling out transits for a GJ~581~e planet with a radius greater than 1.6 R$_{\oplus}$ for most transiting configurations. F2011 rules out transits corresponding to a planet radius greater than 0.75 R$_{\oplus}$, though only within the 1.5$\sigma$ transit window. \citeauthor{dragomir2012b} also excludes transits for GJ~581~b. These results set a limit on the orbital inclination of the system (assuming co-planarity) of $\lesssim$88$^{\circ}$.

Additionally, far-infrared/sub-mm Herschel observations of the GJ~581 constrain the inclination of the system's debris disk to between 27 and 78$^{\circ}$ (1$\sigma$ limits) \citep{lestrade2012}. If the planets are co-planar with the disk, this inclination range suggests planet masses no greater than 2.2 times their radial velocity-measured minimum masses. 

\begin{figure}
\begin{tabular}{ll}
\includegraphics[width=3.25in]{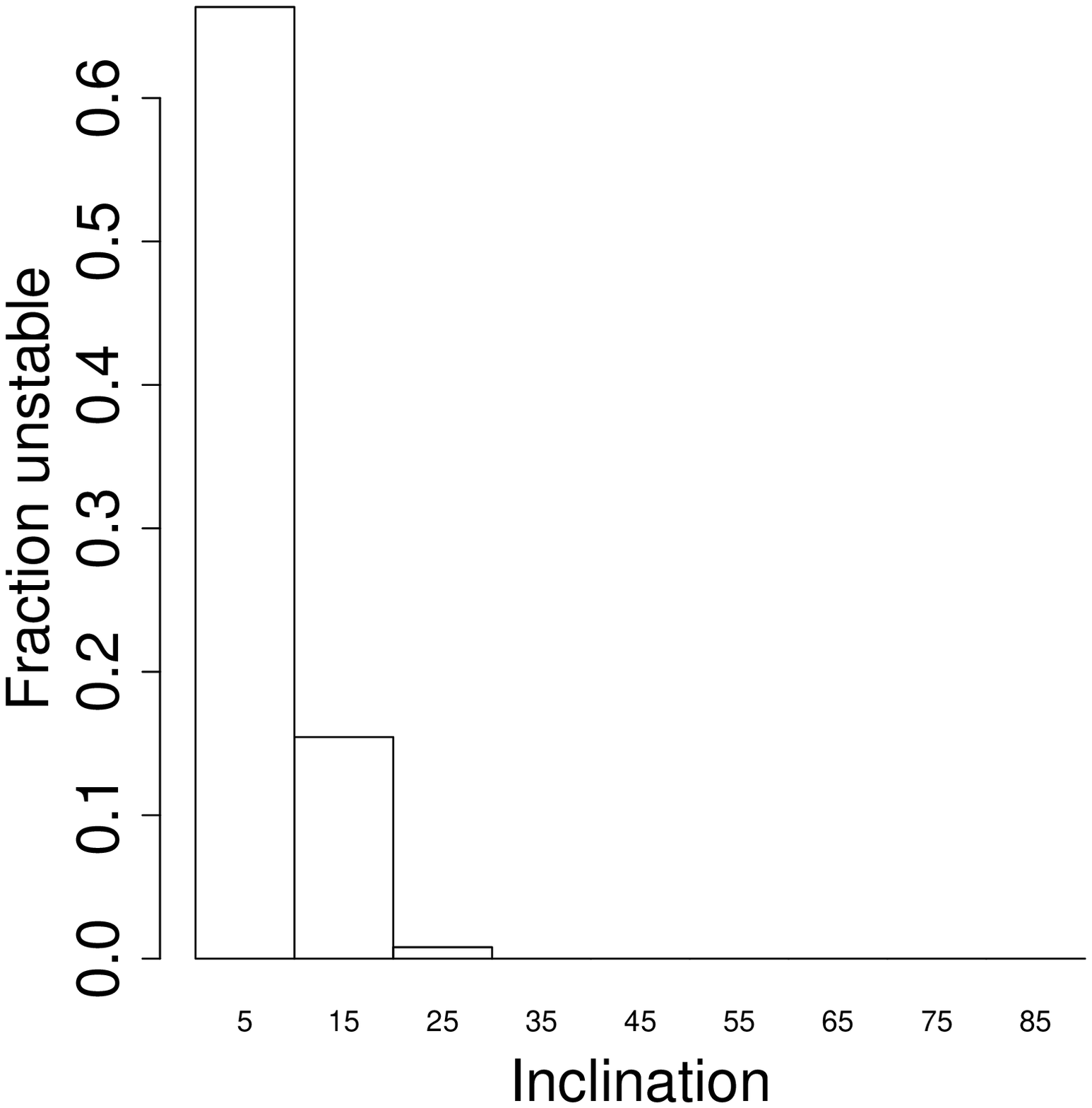}
&
\includegraphics[width=3.25in]{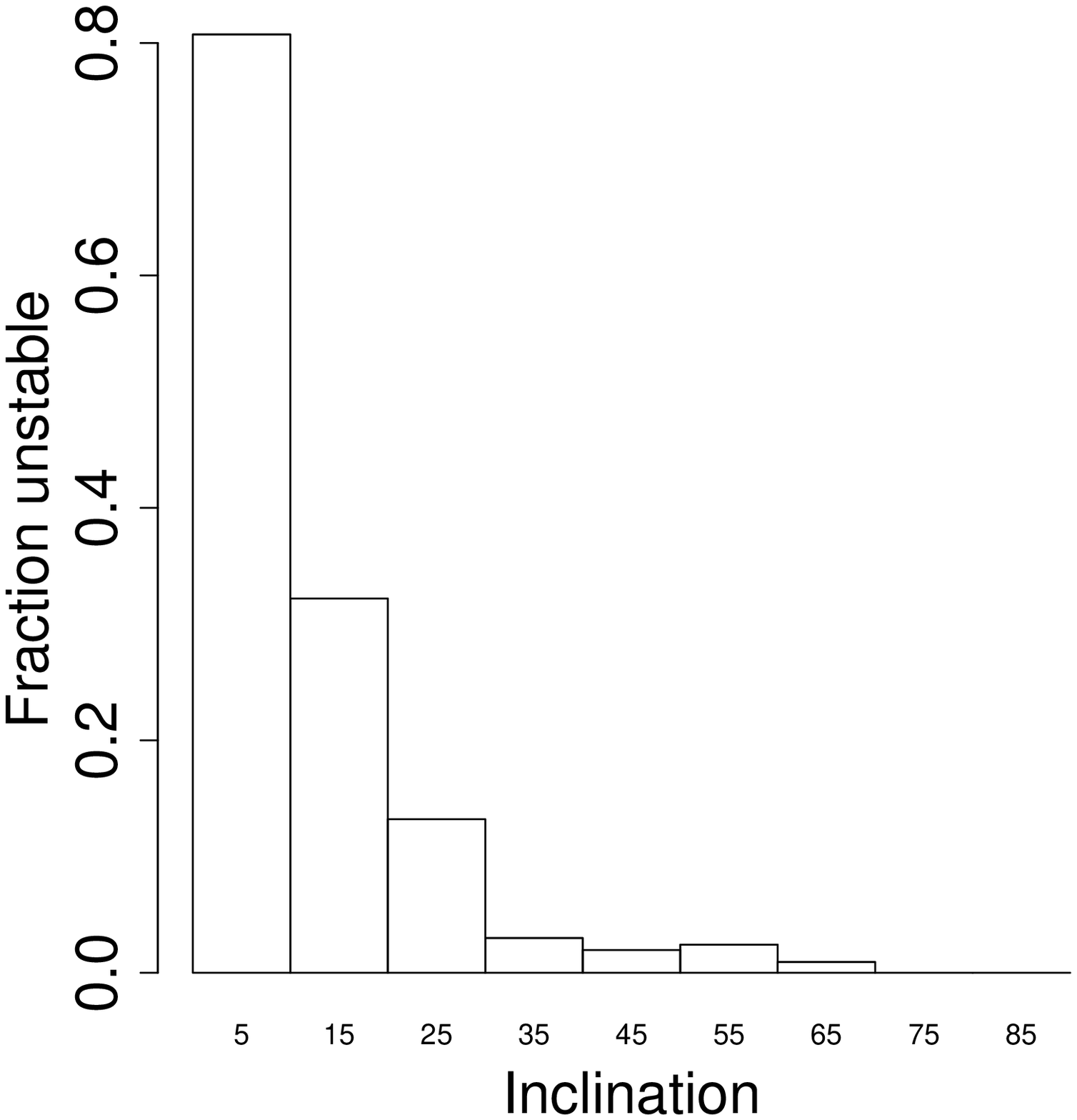}
\\
\includegraphics[width=3.25in]{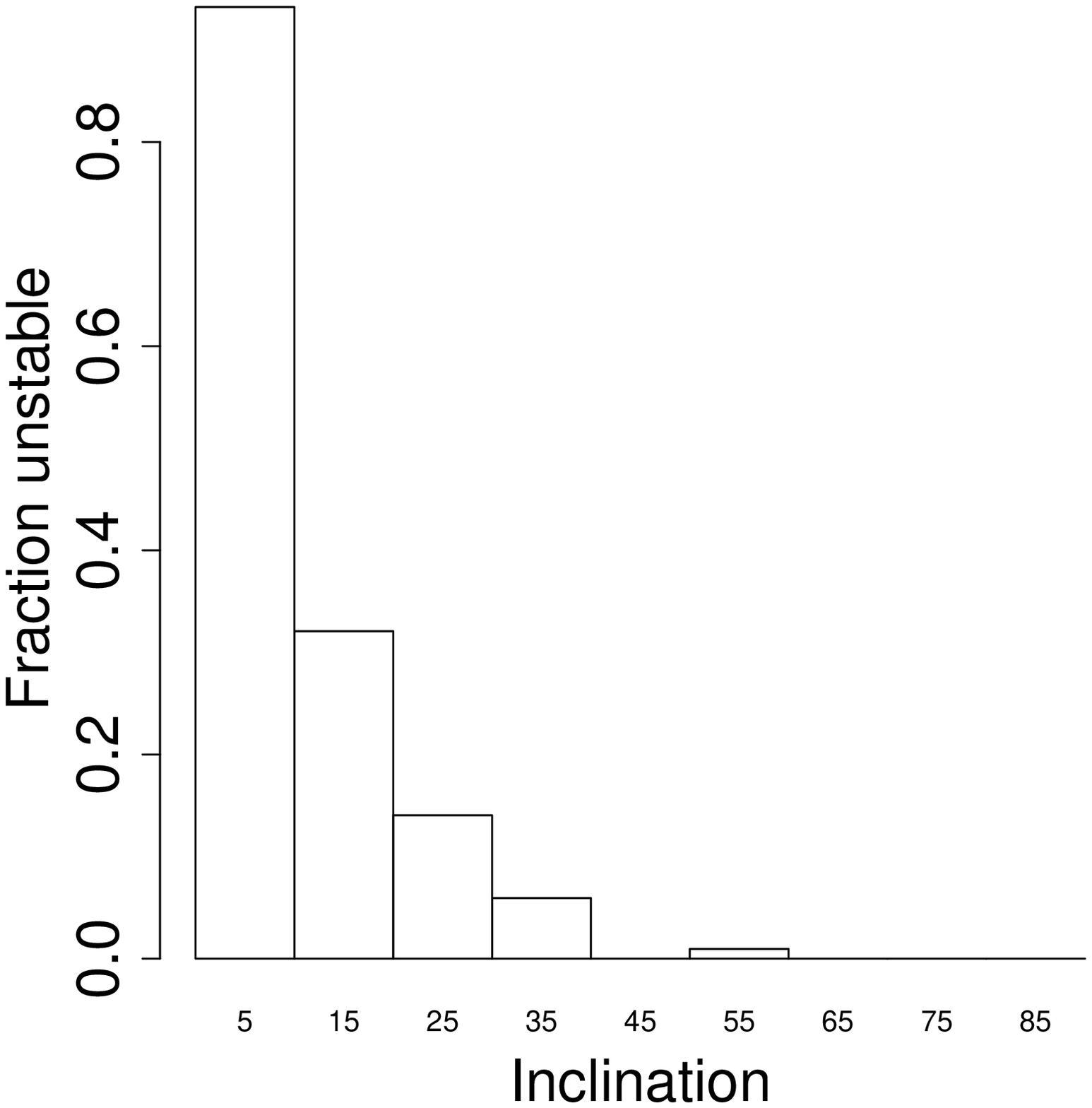}
&
\includegraphics[width=3.25in]{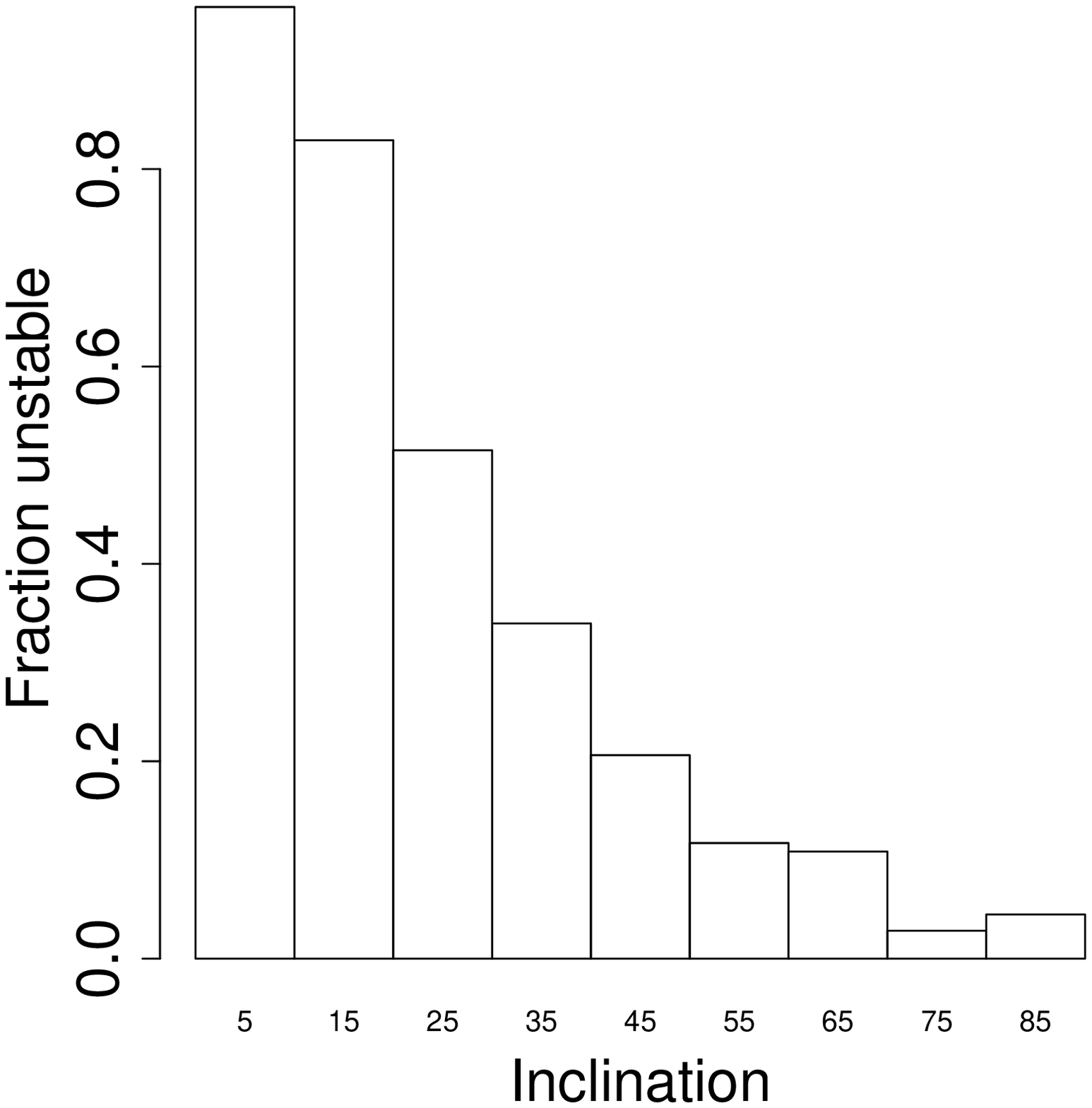}
\\
\end{tabular}
\caption{Fraction of models in a Monte Carlo parameter space study that were unstable within a timescale of $10^6$ years binned by system inclination, assuming a range of stable values for an edge-on system. Top row from right to left shows 4-planet models with and without an additional mutual inclination added to planet. Bottom row shows the same for a 5-planet model. Input into the MC study was taken from the results of Tables \ref{tab:4planetc} and \ref{tab:5planetc}, combined with inclinations chose to cover a uniform distribution from $0$ to $90^\circ$ and additional mutual inclinations chosen uniformly from $-15^\circ$ to $15^\circ$.   \label{fig:inclinations}}
\vspace{0.05in}
\end{figure}

As N-body modeling was performed assuming an edge-on system with no mutual inclination, further models were run over the range of
uncertainties in the fit parameters for both 4 and 5 planet models to determine whether this resulted in any clear limits on the inclination of the system.
1000 trials with random initial conditions were chosen covering a system inclination from 0 to $90^\circ$, based on the ranges of parameters associated with the
``jitter included, stability required'' model fits. Models were classified as stable or unstable and binned by system inclination, with fraction of models in the MC run per bin unstable shown in Figure~\ref{fig:inclinations}. This was repeated with an additional 1000 trials to allow for deviations in each planet's mutual inclination of $\pm 15^\circ$. In each case, a very high system inclination applied to the model fits results in a higher likelihood of instability. The addition of mutual inclination between planets increases this to a small degree, but does not affect the underlying trend. Even in the most extreme case (5-planet with mutual inclination included) $50\%$ if fits that are stable when edge on are still stable at an inclination of $25^\circ$. As can be seen, models remain stable for a significant deviation from an edge-on orientation for both 4 and 5 planet configurations to inclinations well within the range suggested by \citeauthor{lestrade2012}, and inclination constraints alone cannot be used to invalidate either the F2011 or V2012 models of GJ~581.

\section{The Gl 581 Habitable Zone}

A primary aspect of interest in this system is the prospect of
habitable planets. We thus redetermine the location of the planets
with respect to a new calculation of the Habitable Zone (HZ).  The HZ
is defined as the region around a star where water can exist in a
liquid state on the surface of a planet with sufficient atmospheric
pressure. Empirical calculations of the HZ boundaries by \citet{kas93}
have been re-interpreted as analytical functions of stellar effective
temperature and luminosity by several authors including
\citet{sel07}. These have since been replaced by revised calculation
performed by \citet{kop13} which includes an extension of the
methodology to later spectral types. These calculations are available
through the Habitable Zone Gallery \citep{kan12a}, which provides HZ
calculations for all known exoplanetary systems.

\begin{figure}
  \includegraphics[scale=0.5,angle=270]{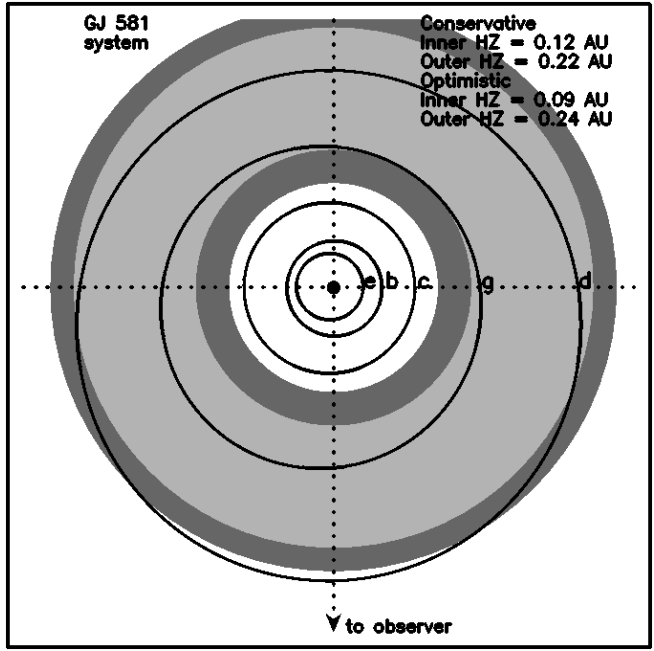}
   \vspace{1.0in}  
\caption{The calculated extent of the conservative (light-gray) and
    optimistic (dark-gray) HZ for the GJ~581 system along with the
    Keplerian orbits of the planets shown in Table
    \ref{tab:5planetc}.}
  \label{hzfig}
\end{figure}

Shown in Figure~\ref{hzfig} is a top-down representation of the GJ~581
system. The orbits of the planets use the Keplerian orbital solutions
shown in Table~\ref{tab:5planetc}. The boundaries of the HZ were
calculated using the stellar parameters of \citet{von11}. The light
gray region represents the ``conservative'' HZ model which uses the
``Runaway Greenhouse'' and ``Maximum Greenhouse'' criteria for the
inner and outer HZ boundaries respectively. The dark gray regions
represent extensions to the HZ (``optimistic model'') which use the
``Recent Venus'' and ``Early Mars'' criteria for the inner/outer HZ
boundaries. These criteria are described in detail by \citet{kan13}
and \citet{kop13}. By these criteria, the g planet, should it be confirmed
by further measurement, spends 100\% of its orbit within the conservative HZ
and would thus be classified as a HZ planet.
The d planet spends 50\% of its orbit within the conservative
HZ and an additional 23\% in the optimistic HZ. The remaining 27\% of
the orbit occurs close to apastron where a significant freezing of
surface liquid water may occur. The ability of the d planet to adapt
to such changes depends strongly upon atmospheric pressure and its
heat distribution efficiency \citep{kan12b}.

\section{Conclusions}  

Using either the fit eccentricities relative to parameter space study determined critical values or N-body model lifetimes relative to some reasonable fraction of a solar lifetime as discrepancy variables, we can test the null hypothesis that the 4- and 5-planet models are stable. The hypothesis that the 4-planet Keplerian model for GJ~581 is stable fails rejection with a p-value of $0.59$ based on stability cutoffs on the results of an RV fit, and fails rejection with $p=0.50$ by sampling the posterior distribution for N-body model lifetime. The hypothesis that the 5-planet (including g) Keplerian model for GJ~581 is stable fails rejection with $p=0.21$ using eccentricity cutoffs and $p=0.10$ by sampling the posterior distribution. Neither the 4- nor 5-planet model can be categorically ruled out on stability grounds.

When the constraint of stability is included in the model fits for 4 and 5 planet Keplerian models of GJ~581, the best fit for a constrained 4-planet model is 2.78 with 21 fit parameters and 240 data points. Similarly, 26 fit parameters for the 5-planet model leads to a best fit reduced chi squared value of 2.58. The F-value of the two distributions is 1.077, and the null hypothesis that they represent the same variance fails rejection with a p-value of .3.   The semi-major amplitude of planet g is about $0.9 m/s$, compared to a mean instrumental error of $1.24 m/s$, a RMS residual of about $2.0 m/s$, and a stellar jitter of $1.4 m/s$.  Previous studies have predicted a stellar jitter due to surface effects of $1.9 m/s$ \citep{vogt2010}. While planet g cannot be ruled out on stability grounds, it does pose greater stability concerns than a 4 planet model, and when stability is a factor in determining best fits to the radial velocity data does not result in enough of an improvement in the fit to pass an F-test. Given that the semi-major amplitude for planet g is lower than the instrumental error, the expected stellar jitter for a star of this type, and the residual noise in the fit, we cannot claim at this time that the data confirms the existence of planet g. The stable fraction of the posterior distribution of  $10\%$ for the 5-planet model does indicate a strong potential for instability in a model for GJ~581 that includes planet g, and any further measurements used to validate the existence of g should include an in depth stability analysis. 

For additional multi-planet systems where stability is a concern, a key orbital parameter affecting stability is orbital eccentricity. One approach to including stability constraints in RV fits could be to start with a parameter space exploration focused primarily on limiting eccentricity, and then once upper bounds for reasonable eccentricities have been defined re-run the RV fit with limits on the eccentricities. This can be efficient, as it allows for a large sampling of the RV fit space, and greater detail in the shape of the posterior distribution; however, it can overestimate the stable portion of the posterior distribution, in particular missing cases where combined effects from multiple planets result in instability at lower eccentricities than from a single planet alone. Starting from the fit result using an MCMC algorithm and then sampling from that distribution as input to N-body simulation provides a better estimate of the likelihood a given planetary model is stable, but requires significantly greater computer time to build a detailed posterior distribution. With either approach, individual N-body simulations require long run-time with few objects and are not easily sped up with parallel computing, but both approaches require many such models to be run, all of which are independent of each other and parallelism can be used to speed up the calculation of the ensemble. Initial tests show that this approach could be applied to other multi-planet systems using modest computational hardware by taking advantage of GPGPU technology.

\acknowledgements

The authors greatly acknowledge the support of the National Science Foundation through grants OCI-0722790, OCI-0959504, and AST-1109662, and of the National Aeronautics and Space Administration through the Astrobiology Institute's Virtual Planetary Laboratory. 

\bibliography{join0214}

\newpage

\newpage

\newpage

\newpage

\newpage

\newpage

\newpage

\newpage

\newpage

\begin{deluxetable}{lcccc}
\tabletypesize{\scriptsize}
\tiny
\tablecaption{4-Planet Model without Jitter \label{tab:4planetnoj}}
\tablehead{
\colhead{Orbital Property} & \colhead{e} &\colhead{b} &\colhead{c} & \colhead{d} 
 } 
\startdata
P (days) & $3.149 \pm 0.1122$ &  $5.369 \pm 5.489e$--$05$ &  $12.92 \pm 0.001522$ &  $66.63 \pm 1.47$ \\
K (m/s) & $1.913 \pm 0.2044$ &  $12.68 \pm 0.1081$ &  $3.174 \pm 0.11$ &  $2.103 \pm 0.2224$ \\
M (degrees) & $332.9 \pm 79.82$ &  $211 \pm 22.94$ &  $138.6 \pm 54.22$ &  $74.1 \pm 55.11$ \\
$e$ & $0.2996 \pm 0.1406$ &  $0.02963 \pm 0.01099$ &  $0.05653 \pm 0.03359$ &  $0.2492 \pm 0.06485$ \\
$\omega$ (degrees) & $172.1 \pm 43.91$ &  $307.8 \pm 38.39$ &  $219.9 \pm 54.04$ &  $261.8 \pm 151.9$ \\
m $\sin I$ ($M_{\odot}$) & $5.744e$--$06 \pm 5.498e$--$07$ &  $4.773e$--$05 \pm 4.077e$--$07$ &  $1.599e$--$05 \pm 5.572e$--$07$ &  $1.775e$--$05 \pm 1.746e$--$06$ \\
a (AU) & $0.02846 \pm 0.0006549$ &  $0.04062 \pm 2.771e$--$07$ &  $0.07293 \pm 5.721e$--$06$ &  $0.2177 \pm 0.003164$ \\
$V_0$ (m/s) & \multicolumn{4}{c}{   -0.388374   }  \\
$\chi_R^2$ & \multicolumn{4}{c}{    2.76403     } \\
RMS (m/s) & \multicolumn{4}{c}{      1.95046     }  \\
Jitter (m/s) & \multicolumn{4}{c}{     1.53488 }
\enddata
\tablecomments{Measure of central tendency is given as the 10 percent trimmed mean of those random walk values with reduced chi-squared within 5 percent of the best value found in the random walk. }
\end{deluxetable}

\newpage

\begin{deluxetable}{lcccc}
\tabletypesize{\scriptsize}
\tiny
\tablecaption{4-Planet Model with Jitter \label{tab:4planet}}
\tablehead{
\colhead{Orbital Property} & \colhead{e} &\colhead{b} &\colhead{c} & \colhead{d} 
 } 
\startdata

P (days) & $3.149 \pm 0.0116$ &  $5.369 \pm 9.413e$--$05$ &  $12.92 \pm 0.002425$ &  $66.65 \pm 1.672$  \\
K (m/s) & $1.854 \pm 0.2196$ &  $12.68 \pm 0.181$ &  $3.165 \pm 0.1842$ &  $2.134 \pm 0.2545$  \\
M (degrees) & $314.1 \pm 131.2$ &  $195.8 \pm 36.16$ &  $164.6 \pm 78.07$ &  $69.42 \pm 56.17$  \\
$e$ & $0.2085 \pm 0.1207$ &  $0.02589 \pm 0.01427$ &  $0.04673 \pm 0.041$ &  $0.2352 \pm 0.09208$  \\
$\omega$ (degrees) & $164.5 \pm 61.46$ &  $318.3 \pm 88.09$ &  $194.8 \pm 78.65$ &  $178.6 \pm 151.6$  \\
m $\sin I$ ($M_{\odot}$)  & $5.691e$--$06 \pm 6.406e$--$07$ &  $4.775e$--$05 \pm 6.816e$--$07$ &  $1.594e$--$05 \pm 9.292e$--$07$ &  $1.805e$--$05 \pm 2.032e$--$06$  \\
a (AU) & $0.02846 \pm 6.898e$--$05$ &  $0.04062 \pm 4.765e$--$07$ &  $0.07293 \pm 9.124e$--$06$ &  $0.2178 \pm 0.003562$  \\
$V_0$ (m/s) & \multicolumn{4}{c}{-0.37104} \\
RMS(m/s) & \multicolumn{4}{c}{1.9541}  
\enddata
\end{deluxetable}

\newpage

\begin{deluxetable}{lcccc}
\tabletypesize{\scriptsize}
\tiny
\tablecaption{4-Planet Model with Jitter and constraints \label{tab:4planetc}}
\tablehead{
\colhead{Orbital Property} & \colhead{e} &\colhead{b} &\colhead{c} & \colhead{d} 
 } 
\startdata
P (days)     & $3.14 9 \pm 0.03428$ &  $5.369 \pm 9.171e$--$05$ &  $12.92 \pm 0.00236$ &  $66.65 \pm 5.87$  \\
K (m/s)    & $1.807 \pm 0.2262$ &  $12.72 \pm 0.1791$ &  $3.155 \pm 0.1898$ &  $2.096 \pm 0.2829$  \\
M (degrees)   & $248.1 \pm 109.8$ &  $191.4 \pm 34.33$ &  $152.7 \pm 83.85$ &  $67.81 \pm 63.41$  \\
$e$ & $0.1021 \pm 0.0634$ &  $0.02365 \pm 0.01362$ &  $0.04476 \pm 0.04087$ &  $0.2227 \pm 0.0971$  \\
$\omega$ (degrees)    & $168.9 \pm 86.65$ &  $322.1 \pm 87.52$ &  $207.6 \pm 84.1$ &  $149.3 \pm 150.5$  \\
m $\sin I$ ($M_{\odot}$)   & $5.658e$--$06 \pm 7.056e$--$07$ &  $4.789e$--$05 \pm 6.759e$--$07$ &  $1.59e$--$05 \pm 9.567e$--$07$ &  $1.776e$--$05 \pm 2.259e$--$06$  \\
a (AU)    & $0.02846 \pm 0.0002025$ &  $0.04062 \pm 4.65e$--$07$ &  $0.07293 \pm 8.871e$--$06$ &  $0.2178 \pm 0.01119$  \\
$V_0$     & \multicolumn{4}{c}{      -0.457253   } \\
RMS (m/s) & \multicolumn{4}{c}{    1.95341     }
\enddata
\end{deluxetable}

\newpage

\begin{deluxetable}{lccccc}
\rotate
\tablewidth{6.75in}
\tabletypesize{\scriptsize}
\tiny
\tablecaption{5-Planet Model without Jitter \label{tab:5planetnoj}}
\tablehead{
\colhead{Orb Prop} & \colhead{e} &\colhead{b} &\colhead{c} & \colhead{g} & \colhead{d} 
 } 
\startdata
P (days) & $3.15 \pm 0.009288$ &  $5.369 \pm 5.906e$--$05$ &  $12.92 \pm 0.001798$ &  $40.55 \pm 2.943$ &  $66.6 \pm 2.286$  \\
K (m/s) & $1.872 \pm 0.1883$ &  $12.66 \pm 0.1155$ &  $3.275 \pm 0.1368$ &  $1.704 \pm 0.4204$ &  $2.168 \pm 0.1916$  \\
M (degrees) & $344.3 \pm 108.8$ &  $215.7 \pm 30.64$ &  $135.9 \pm 55.9$ &  $113.7 \pm 90.61$ &  $86.67 \pm 41.26$  \\
$e$ & $0.296 \pm 0.117$ &  $0.03301 \pm 0.01097$ &  $0.07286 \pm 0.03829$ &  $0.864 \pm 0.2677$ &  $0.215 \pm 0.1052$  \\
$\omega$ (degrees)  & $167.2 \pm 37.05$ &  $302.7 \pm 41.71$ &  $229.6 \pm 56.09$ &  $254.5 \pm 83.02$ &  $331.4 \pm 147.8$  \\
m $\sin I$ ($M_{\odot}$)  & $5.64e$--$06 \pm 4.34e$--$07$ &  $4.77e$--$05 \pm 4.35e$--$07$ &  $1.65e$--$05 \pm 6.95e$--$07$ &  $6.34e$--$06 \pm 1.10e$--$06$ &  $1.85e$--$05 \pm 1.83e$--$06$  \\
a (AU)  & $0.02846 \pm 5.593e$--$05$ &  $0.04062 \pm 2.999e$--$07$ &  $0.07293 \pm 6.757e$--$06$ &  $0.1563 \pm 0.008056$ &  $0.2177 \pm 0.004893$  \\
$V_0$ (m/s)     & \multicolumn{5}{c}{  -0.505305 } \\
$\chi_R^2$      & \multicolumn{5}{c}{    2.48034 } \\
RMS (m/s)        & \multicolumn{5}{c}{      1.85163 } \\
jitter (m/s)      & \multicolumn{5}{c}{   1.42592 } 
\enddata
\end{deluxetable}

\newpage

\begin{deluxetable}{lccccc}
\rotate
\tablewidth{7in}
\tabletypesize{\scriptsize}
\tiny
\tablecaption{5-Planet Model with Jitter \label{tab:5planet}}
\tablehead{
\colhead{Orbital Property} & \colhead{e} &\colhead{b} &\colhead{c} & \colhead{g} & \colhead{d} 
 } 
\startdata
P (days)  & $3.149 \pm 0.2583$ &  $5.369 \pm 6.686e$--$05$ &  $12.92 \pm 0.001931$ &  $34.45 \pm 3.218$ &  $66.61 \pm 2.006$  \\
K (m/s) & $1.853 \pm 0.1563$ &  $12.69 \pm 0.128$ &  $3.175 \pm 0.1426$ &  $1.242 \pm 0.4444$ &  $2.21 \pm 0.195$  \\
M (degrees) & $284.1 \pm 147.5$ &  $196.4 \pm 27.17$ &  $158.1 \pm 57.18$ &  $189.2 \pm 101.7$ &  $84.76 \pm 47.03$  \\
$e$ & $0.2014 \pm 0.1044$ &  $0.0268 \pm 0.01135$ &  $0.05571 \pm 0.03613$ &  $0.5988 \pm 0.2652$ &  $0.1728 \pm 0.09115$  \\
$\omega$ (degrees)  & $158.4 \pm 45.36$ &  $319.5 \pm 80.7$ &  $206.7 \pm 57.91$ &  $215.2 \pm 111.9$ &  $328 \pm 131.4$  \\
m $\sin I$ ($M_{\odot}$)  & $5.71e$--$06 \pm 4.62e$--$07$ &  $4.78e$--$05 \pm 4.82e$--$07$ &  $1.60e$--$05 \pm 7.20e$--$07$ &  $6.23e$--$06 \pm 1.45e$--$06$ &  $1.90e$--$05 \pm 1.66e$--$06$  \\
a (AU)  & $0.02846 \pm 0.001371$ &  $0.04062 \pm 3.38e$--$07$ &  $0.07293 \pm 7.259e$--$06$ &  $0.1401 \pm 0.008629$ &  $0.2177 \pm 0.004255$  \\
$V_0$ (m/s)     & \multicolumn{5}{c}{  -0.385183 } \\
RMS (m/s)        & \multicolumn{5}{c}{      1.83831 } 
\enddata
\end{deluxetable}

\newpage

\begin{deluxetable}{lccccc}
\rotate
\tablewidth{7in}
\tabletypesize{\scriptsize}
\tiny
\tablecaption{5-Planet Model with Jitter and constraints \label{tab:5planetc}}
\tablehead{
\colhead{Orbital Property} & \colhead{e} &\colhead{b} &\colhead{c} & \colhead{g} & \colhead{d} 
 } 
\startdata
P (days) & $3.149 \pm 0.000274$ &  $5.369 \pm 9.511e$--$05$ &  $12.92 \pm 0.002721$ &  $32.05 \pm 5.049$ &  $66.61 \pm 2.558$  \\
K (m/s) & $1.816 \pm 0.2114$ &  $12.65 \pm 0.1808$ &  $3.125 \pm 0.1766$ &  $0.7945 \pm 0.3327$ &  $2.2 \pm 0.2572$  \\
M (degrees) & $261.1 \pm 123.1$ &  $199.6 \pm 23.6$ &  $146.6 \pm 76.97$ &  $180.9 \pm 100.5$ &  $88.61 \pm 76.48$   \\
$e$ & $0.1146 \pm 0.05816$ &  $0.02766 \pm 0.01333$ &  $0.05265 \pm 0.04694$ &  $0.133 \pm 0.07239$ &  $0.1268 \pm 0.07039$   \\
$\omega$ (degrees)  & $168.8 \pm 88.03$ &  $318.9 \pm 32.34$ &  $206.6 \pm 80.36$ &  $239.4 \pm 104.7$ &  $277.2 \pm 134.5$   \\
m $\sin I$ ($M_{\odot}$)  & $5.68e$--$06 \pm 6.57e$--$07$ &  $4.76e$--$05 \pm 6.84e$--$07$ &  $1.57e$--$05 \pm 8.98e$--$07$ &  $5.42e$--$06 \pm 2.27e$--$06$ &  $1.90e$--$05 \pm 2.10e$--$06$  \\
a (AU)  & $0.02846 \pm 1.651e$--$06$ &  $0.04062 \pm 4.805e$--$07$ &  $0.07293 \pm 1.025e$--$05$ &  $0.1336 \pm 0.01343$ &  $0.2177 \pm 0.00541$  \\
$V_0$ (m/s)     & \multicolumn{5}{c}{  -0.397863 } \\
RMS (m/s)        & \multicolumn{5}{c}{      1.88801 }
\enddata
\end{deluxetable}

\end{document}